\documentclass{IEEEtaes}

\usepackage{color,array,amsthm}
\usepackage{graphicx}
\usepackage[utf8]{inputenc}
\usepackage[T1]{fontenc}
\usepackage{orcidlink}
\usepackage{cite}
\usepackage{amsmath,amssymb,amsfonts}

\usepackage{algorithm}
\usepackage{algpseudocode}
\usepackage{graphicx}
\usepackage{textcomp}
\usepackage{textcomp}
\usepackage{xcolor}
\usepackage{placeins}
\usepackage{makecell}
\usepackage{float}
\usepackage{subcaption}

\usepackage{jabbrv}
\DefineSpuriousJournalWord{on}
\DefineSpuriousJournalWord{the}
\DefineSpuriousJournalWord{of}

\usepackage{url}
\usepackage[nameinlink]{cleveref}
\crefname{equation}{}{}
\crefname{section}{Sec.}{Secs.}
\crefname{figure}{Fig.}{Figs.}
\crefname{table}{Table}{Tables}

\jvol{62}
\jnum{XX}
\jmonth{XXXXX}
\paper{1234567}
\pubyear{2026}
\doiinfo{TAES.2026.Doi Number}

\begin{document}

\title{Deep Learning-Based Detection of Electrical Faults and Power Quality Disturbances in Aerospace Power Systems}

\author{Ian C. Guzman\orcidlink{0000-0002-4532-8814}}
\affil{Embry-Riddle Aeronautical University, Daytona Beach, FL 32114, USA} 

\author{Radu Babiceanu\orcidlink{0000-0003-0705-1452}}
\member{Senior Member, IEEE}
\affil{Western Michigan University, Kalamazoo, MI 49008, USA} 

\author{Berker Peköz\orcidlink{0000-0002-7572-3663}}
\member{Member, IEEE}
\affil{Embry-Riddle Aeronautical University, Daytona Beach, FL 32114, USA} 

%% \author{FOURTH D. AUTHOR}
%% \affil{University of Colorado, Colorado, USA}

\receiveddate{
Manuscript received December 4, 2025; revised July 10, 2026; accepted September 2, 2026.\\
% This paragraph of the first footnote will contain the date on which you submitted your paper for review, which is populated by IEEE. It is IEEE style to display support information, including sponsor and financial support acknowledgment, here and not in an acknowledgment section at the end of the article. For example, ``This work was supported in part by the U.S. Department of Commerce under Grant BS123456.'' 
}
%% \accepteddate{XXXXX XX XXXX}
%% \publisheddate{XXXXX XX XXXX}

\corresp{{\itshape (Corresponding author: I. Guzman)} (e-mail: \href{mailto:guzmani@my.erau.edu}{guzmani@my.erau.edu}).}

\authoraddress{}

\editor{}
% {Mentions of supplemental materials and animal/human rights statements can be included here.}
\supplementary{Color versions of one or more of the figures in this article are available online at \href{http://ieeexplore.ieee.org}{http://ieeexplore.ieee.org}.}

\markboth{GUZMAN ET AL.}{DL-Based Detection of Electrical Faults and PQDs in Aerospace Power Systems}
\maketitle

\begin{abstract} The transition toward More Electric Aircraft (MEA) has introduced highly complex electrical architectures that impose strict requirements on reliability, safety, and real‑time operation. Yet most existing research on power quality disturbances (PQDs) and electrical fault diagnosis targets conventional utility‑scale power grids and relies on low‑frequency analysis, which limits accuracy and applicability in aircraft electrical systems that operate at higher frequencies. As a result, the use of data‑driven PQD classification in aircraft power networks remains largely unexplored. This paper addresses this gap by presenting a deep learning–based framework for automated multiclass detection and classification of electrical faults and PQDs in aircraft electrical systems with emphasis on classification metrics, robustness, and applicability under aerospace constraints. A high‑fidelity aircraft power system model inspired by the Boeing 787 electrical architecture was developed to represent operation at a 400Hz fundamental frequency. The model produced high‑resolution signals under a comprehensive set of fault and PQD conditions. Two datasets were produced. The first dataset consists of one‑dimensional time‑domain signals, further augmented through signal processing techniques and generative adversarial networks (GANs) to increase data diversity and robustness; this dataset is made publicly available on IEEE DataPort and supports the evaluation of one‑dimensional convolutional neural network (1D‑CNN) models. The second dataset consists of two‑dimensional time–frequency representations derived from the short‑time Fourier transform and targets two‑dimensional convolutional architectures (2D-CNN). Several deep learning architectures were evaluated, including 1D‑CNNs, 2D-CNNs, LSTMs, hybrid CNN–LSTM models, and established deep convolutional networks such as ResNet, MobileNet, and VGG. A compact ResNet architecture demonstrated the most favorable balance between classification performance and model complexity, achieving a software test accuracy of 96.94\% with only 175,685 parameters. After 8-bit quantization and deployment on a Xilinx Zynq UltraScale+ MPSoC ZCU102, the FPGA accelerator achieved a post-quantized accuracy of 95.87\%, with a mean on-board inference latency of 6.90 ms and a worst-case on-board latency of 14.40 ms per classifier input record. The implementation used 15.52\% of DSP slices, 63.71\% of Block RAM, and 93.77\% of CLB LUTs, indicating real-time feasibility for the deep-learning inference accelerator while leaving limited LUT margin for additional on-FPGA integration. The findings show that lightweight deep learning models can provide accurate multiclass classification under aerospace-relevant resource and timing constraints.

\end{abstract}

\begin{IEEEkeywords}
aircraft power systems, fault diagnosis, field programmable gate arrays (FPGAs), neural networks, power quality, power system faults.
% Aerospace power systems, deep learning, electrical fault detection, FPGA, More Electric Aircraft, power quality disturbances, residual networks, real-time inference.
% Enter keywords or phrases in alphabetical order, separated by commas. For a list of suggested keywords, send a blank e-mail to \href{mailto:keywords@ieee.org}{keywords@ieee.org} or visit \href{http://www.ieee.org/organizations/pubs/ani\_prod/keywrd98.txt}{\url{http://www.ieee.org/organizations/pubs/ani\_prod/keywrd98.txt}}
\end{IEEEkeywords}

\section{Introduction}
T{\scshape he} aerospace industry is transitioning toward More Electric Aircraft (MEA); replacing conventional hydraulic, pneumatic, and mechanical components with advanced electrical systems\cite{Toward_more_electric_powertrains_in_aircraft}; to reduce operational expenses such as fuel, emission and maintenance costs while improving efficiency, system reliability and performance\cite{Analysis_PQ_aviation,Insulation_materials}. While these significant benefits led to growing adoption, this shift requires complex electrical and electronics networks operating at higher voltages and frequencies, creating  unprecedented complexity and reliability challenges in power system design, implementation, maintenance, and fault management  \cite{More_Electric_Aircraft:_Review_Challenges, Switched-Reluctance_Motor_Drive_for_MEA}.

For example, higher operational voltages bring significant challenges for aerospace insulation systems as the risk of degradation from partial discharge (PD) increases\cite{Arc_faults_ref47}. At low pressures, electrical insulation systems become more vulnerable to PDs, while the harsh environments and confined spaces in aircraft further complicate insulation reliability \cite{Arcfaults}. Electrical systems intended for ground operations can experience PDs in their insulation at flight altitude due to low-pressure conditions. These discharges become more severe as moisture condenses following sudden pressure changes during rapid ascent and descent, underscoring the need for advanced monitoring solutions capable of detecting and classifying electrical faults real-time under stringent operational constraints\cite{Arc_faults_ref7}. This increased complexity further introduces challenges in maintaining optimal power quality (PQ), as PQ disturbances (PQDs) affect not only the power grid and residential electrical systems but also the aerospace industry\cite{Analysis_PQ_aviation}. PQDs are caused by non-linear loads, including power converters, variable speed drives, avionics, entertainment and communication systems, and incandescent and LED lighting technologies. PQDs lead to operational disruptions, reduced efficiency, equipment malfunctions, or even permanent damage to critical onboard systems; compromising flight safety.

Implementing continuous power monitoring systems in aircraft is essential to ensure the safe, reliable, and efficient MEAs. Early detection of electrical faults such as short circuits and open circuits allows issues to be addressed before they escalate into significant failures, thus preventing disruptions and minimizing downtime. This proactive predictive maintenance approach extends lifespan of critical components, leading to improved aircraft availability and operational performance. As electrification technologies continue to be adopted by the aerospace sector, the deployment of power monitoring systems will remain indispensable to guarantee the stable, safe, and efficient operation of increasingly complex aircraft electrical systems.

% The contribution of this work is the development of ML-based methods tailored to the classification of faults and PQDs in aircraft electrical systems. 
While there is extensive research on classifying PQDs in residential electrical systems using ML models, there is a lack of studies specifically focused on aircraft power systems. The unique characteristics of aircraft electrical networks such as higher operation frequencies than traditional electrical systems and compact system architecture present different challenges that are not typically addressed in conventional power system studies.
Additionally, the critical nature of aircraft power systems where electrical disturbances can have serious safety and operational implications further reinforces the need for specialized approaches. Current deep learning (DL) models and techniques used for residential or industrial power systems are not directly applicable due to the differences in system dynamics, environmental factors, and real-time operational constraints. %To address these challenges and the research gap in the existing literature, this research focuses on developing DL-based methods to classify faults and PQDs in aircraft power systems taking into account the specific operational conditions, frequency ranges, and reliability requirements unique to the aerospace industry.

To bridge these gaps, this work proposes a DL-based framework for automated detection and multiclass classification of electrical faults and PQDs in aerospace power systems. The approach leverages high-fidelity signal generation, data augmentation, and generative adversarial networks (GANs) to overcome data scarcity. Multiple architectures, including CNNs, LSTMs, and hybrid models, are evaluated for accuracy and computational efficiency. A compact residual network is introduced to achieve near state-of-the-art performance with minimal parameters, enabling deployment on resource-constrained platforms.
The contributions of this paper are threefold:
\begin{enumerate}
    \item Generating a comprehensive dataset spanning faults and PQDs suitable for DL training purposes using high-fidelity simulation model inspired by the Boeing 787 electrical architecture,
    \item 
Design and evaluation of lightweight DL models for real-time fault \& PQD classification, and
\item FPGA-based implementation demonstrating efficient resource utilization and compliance with aerospace hardware constraints.
\end{enumerate}

These findings advance intelligent health monitoring for MEA systems, supporting predictive maintenance and enhancing operational reliability in next-generation aircraft. 
The rest of this article is organized as follows: related work on aircraft electrical faults, PQDs in aircraft electrical systems and FPGA-based electrical fault/PQD detection DL implementations are surveyed in \cref{sec:relw}, the electrical model used in this work is detailed in \cref{sec:model}, the software experiments and results are explored in \cref{sec:exp}, the FPGA implementation and results are described in \cref{sec:impl}, and finally, \cref{sec:conc} discusses conclusions and outlines directions for future research.

\section{Related Work\label{sec:relw}}
\subsection{Electrical faults in aircraft power systems}
Most of the current research on fault diagnosis of aerospace power systems is based on steady-state signals or low frequency approaches. For instance, in \cite{ref13} the authors classified electrical failures using the steady state signals by simulating a number of fault conditions of the Boeing 787 power system. In \cite{RefSELS} the authors classified electronic load faults in spacecraft power systems using their private dataset that they called "SELS" using convolutional neural networks. In \cite{refharm} the authors used harmonics to localize faults in aircraft power systems. A spacecraft simulation system developed by NASA called "Virtual Adapt" for research in fault diagnosis of aerospace power systems is a high-fidelity MATLAB-Simulink model of the Advanced Diagnostic and Prognostic Testbed (ADAPT) deployed at NASA Ames Research Center and it includes: power generation, storage and distribution systems. This resource can be found online as a public resource and it's described in \cite{RefVA} but it's very limited as it was largely modeled with direct current rather than alternating current and hence it only allows to run simulations for DC signals. The authors in \cite{Ref_estDC} estimated faults in DC power systems by means of mathematical models using the Virtual Adapt model. However, low-frequency methods do not provide enough information to distinguish between a large number of electrical loads \cite{refnilm3}. In contrast, high-frequency approaches provide significantly more detailed information about the waveform of each load \cite{refnilm4,refnilm5}, leading to better discrimination and improved load classification \cite{deepDFML}, particularly when using spectral features \cite{refnilm7}. In addition, faults can be detected faster by processing transients instead of steady state signals. 

Traditional data-driven fault diagnosis methods for power systems in general involve two key steps: feature extraction and fault classification. For this approach, the effectiveness of classification largely depends on the feature extraction process which can be both time-consuming and labor-intensive \cite{Progressive_Improved_Convolutional_Neural_Network_for_Avionics}. Features of signals in the time, frequency, and time–frequency domains have been analyzed using signal processing methods and utilized for fault diagnosis along with machine learning (ML)-based classifiers such as support vector machines (SVM), multilayer perceptrons (MLP), k-nearest neighbor (KNN), random forest (RF) \cite{Fault_Diagnosis_for_Rotating_Machinery,fault_diagnosis_random_forest}. However, these approaches require feature engineering which introduces additional challenges such as the need for expert domain knowledge, potential loss of discriminative information \cite{A_Review_of_Feature_Selection_and_Feature_Extraction} and increased computational cost.\\
In recent years, with the development of DL algorithms such as convolutional neural networks (CNN) and recurrent neural networks (RNN) the difficulties and challenges of feature extraction methods from signal-based data have been overcome \cite{Early_Fault_Detection} as these models can process one-dimensional signals such as electrical signals and extract representative features from raw data using deep structures. CNNs perform well at recognizing spatial patterns in images and sounds while RNNs are more effective at processing sequential data, which make them well-suited for time series analysis \cite{Enhanced_Fault_Diagnosis_Lenet}.
%\IEEEpubidadjcol

With respect to aerospace power systems several research works from academia and industry have reported their investigations on fault diagnosis. For instance, according to \cite{Open-circuit_fault_diagnosis_phase_voltage} current diagnostic methods to detect open-circuit faults in the aerospace industry are ineffective, so a method based on phase voltage and control techniques is proposed in that work. In \cite{Fault_diagnosis_current_waveforms} short and open circuits in aircraft power systems are detected using current waveforms and mathematical methods but without the aid of ML algorithms. On the other hand, in \cite{Arc_Fault_Classification_wavelet_GA-RF} research on arc faults in aircraft electrical systems using signal processing methods based on the empirical wavelet transform (EWT) for feature extraction along with genetic algorithms (GA) and RF classifier is performed. Similarly, short and open circuits in aircraft power generation systems are classified in \cite{fault_diagnosis_ISSA_SVM} using GA based on improved squirrel search algorithm (ISSA) along with SVM (ISSA-SVM). In \cite{Power_fault_prediction_datamining} power faults in aircraft are detected by collecting data but relying only on data mining techniques. The authors in \cite{fault_diagnosis_aircraft_petrinets} proposed mathematical models based on Petri Nets to classify faults in distribution lines of aircraft electrical systems. In a recent study \cite{fault_prediction_LSTMSA} on fault detection in aircraft power systems, several ML methods were tested including: gated recurrent units, SVMs, support vector regressors, bidirectional long-short term memory (LSTM). Among these, a classifier called LSTM-Self-Attention (LSTM-SA) achieved better results than the other tested classifiers with an accuracy of 86.61\%. However, this accuracy is relatively low considering the high accuracy required in safety-critical environments such as aerospace systems. Meanwhile, in \cite{Progressive_Improved_Convolutional_Neural_Network_for_Avionics} a stochastic discrete-time series CNN (SDCNN) is proposed for fault diagnosis in aircraft electrical systems.

While there has been some progress in recent years regarding detection of electrical faults in aircraft power systems using ML methods, several challenges still remain including the need to explore more data-driven models that can reach high classification accuracies despite a high number of classes and their real-time deployment in safety-critical environments. Therefore, in order to improve the reliability and effectiveness of fault detection systems further exploration of new approaches, faults, electrical disturbances, architectures, and optimization techniques is required. As these challenges are addressed, DL approaches and their implementation are expected to significantly improve fault detection accuracy and system performance in the future. Thus, research on the detection of electrical faults in aircraft power systems using DL methods is still in its early stages and requires continued advancement and development.

\subsection{PQDs in aircraft power systems}

As the integration rate of non-linear loads increase in modern electrical grids the vulnerability of power systems to PQDs escalates leading to increasing challenges in maintaining stability and reliability \cite{Ravi2023}. These non-linear loads introduce complexities to the grid dynamics characterized by their intermittent, fluctuating and non-stationary nature. Consequently, the likelihood of PQDs being introduced into the power signal rises causing significant risks to the normal operation of the electrical grid. The repercussions on electrical equipment and operational processes include: misoperation, damage, process interruptions, and other irregularities that entail significant costs \cite{ieee1159}. The appearance of PQDs also affects the aerospace industry since over the last few decades there has been an increasingly demand in aircraft electrical systems as a transition trend from the classic fuel-based propulsion designs towards MEA. Non-linear loads in aircraft systems include: power converters such as inverters and rectifiers, avionics, entertainment systems, communication devices, lighting systems (e.g. incandescent and LED), among others. PQ events can affect the quality, reliability, and safety of the aircraft electrical system. Deploying robust PQ monitoring systems in aircraft enables the prevention of the impact of voltage, current and frequency fluctuations by providing insights into their behaviour which help in mitigating PQDs to ensure the reliable operation of aircraft electrical systems maintaining the safety of passengers and crew.

The vast majority of research works have investigated signal processing techniques and DL models for PQDs using the typical residential power system frequency of 50Hz-60Hz whereas aircraft electrical systems usually work at a frequency of 400Hz meaning that signals in aircraft power systems have higher frequency components than signals in residential power grids which indicates that there may be differences between PQDs for commercial grid and aircraft power system signals in the time domain such as variations in the shape, duration or amplitude.

Recent PQD classification studies increasingly use CNN, LSTM, and
hybrid CNN--LSTM architectures applied to one-dimensional time-series
signals or two-dimensional time--frequency representations
\cite{garcia2020pqd,chiam2023pqd}. In parallel, GAN-based approaches
have been explored to increase data diversity for PQD classification under
limited labeled data \cite{jian2021ganpqd}. However, these studies
primarily target conventional utility or renewable-energy power systems.
The use of compact deep learning models for multiclass fault and PQD
classification in 400-Hz aircraft electrical systems, together with
FPGA-level deployment analysis, remains comparatively less explored.

Signal processing techniques are employed in conjunction with AI methods to detect and classify PQDs. Some of the most used signal processing algorithms for feature extraction of PQDs are based on the Wavelet, Fourier and Stockwell transforms \cite{A_Novel_three_step_classification}. Classical signal processing methods such as the Short-Time Fourier Transform (STFT) and the continuous wavelet transform (CWT) cannot achieve high resolution simultaneously in both the time and frequency domains and as a result, classical time-frequency (TF) methods produce blurred TF representations, leading to inaccurate characterization of TF features in non-stationary signals \cite{Synchrosqueezing_Transform,Stockwell}. Some of the most recent used signal processing techniques for time-frequency decomposition of PQDs include Wavelet-based methods. For instance, a method for classifying PQDs using the discrete wavelet transform (DWT) in conjunction with a cubic multi-class SVM (CMSVM) was proposed recently in \cite{Distribution_network_PQ}. The empirical wavelet transform (EWT) was employed recently in \cite{Wavelet_SVM} alongside with a SVM. A Fourier-based method called the Synchrosqueezing Transform was reported in \cite{Synchrosqueezing_Transform} where the authors proposed a methodology to extract non-stationary components from power systems. In \cite{Stockwell} the Stockwell transform was used to extract features from PQDs to be classified with ML models. These signal processing methods require feature engineering, which can be labor-intensive and may not capture all relevant information present in the data. However, DL models can automatically learn and extract patterns and features directly from the raw signal data. This is why signal processing methods based on DL models, like RNN, LSTM  and CNN have gained substantial interest over the last few years in the realm of PQDs \cite{DL7,Wiener_DL,DL2,DL3,DL4,DL5,DL6}. Due to their capacity of processing time series signals and images by extracting features automatically without the need to implement a feature extraction stage, these models can effectively capture complex temporal dependencies and spatial patterns present in PQ data, leading to more accurate detection, classification, and prediction of PQDs. Moreover, the elimination of a separate feature extraction stage simplifies the overall signal processing pipeline, reducing the time and effort required for development and deployment.

\subsection{FPGA-based DL implementation}

Implementing real-time fault management systems on aircraft enhances safety and reliability. These real-time fault management systems can be deployed using FPGAs as they can process large amounts of data in real-time at high speeds. Additionally, FPGAs offer remarkable signal processing-based computational capabilities, compact size, and reduced power consumption which make them appealing solutions in the aerospace industry \cite{fpga-based_military,Airwrothiness_fpga}.

% FPGAs offer a versatile and highly configurable platform ideally suited for real-time processing tasks making them particularly appealing for PQ monitoring applications. For instance, s
Signal processing methods such as the FFT have been implemented on FPGA to classify PQ events as demonstrated in a study reported in \cite{FPGA_biletskiy}. In this study an FFT IP core provided by Xilinx ISE was used to compute the spectrum of PQ events which were then compared with stored spectrums in a PQ database for classification. The DWT was computed in \cite{FPGA_garcia_rodriquez} for feature extraction of PQ events where the authors used a DWT IP core to deploy this algorithm on FPGA. Similarly, the authors in \cite{PQ_labview_fpga} implemented the DWT on FPGA utilizing LabView and stored the obtained wavelet features in a database for classification. In \cite{Hilbert_filter_based_FPGA} the authors designed Hilbert filter banks and comparators on FPGA to classify PQDs within their own experimental setup. The setup included three non-linear loads (two motors and a capacitor bank) that were switched to create real-time PQDs for testing their implementation. In \cite{cubic_spline} the authors proposed an FPGA-based technique for detecting and measuring the time span of voltage and current swells utilizing cubic splines for calculating the signal upper envelope.

Research on PQDs that rely solely on traditional signal processing hardware implementations and/or databases such as the aforementioned works or similar approaches face significant limitations in the classification stage due to the absence of ML-based classifiers or models that can learn from data. On the other hand, FPGAs have become a very attractive reconfigurable platform for ML implementations since they offer parallel processing capabilities enabling simultaneous execution of multiple computations. This aligns well with the inherently parallel nature of many ML algorithms such as neural networks. In recent years, a limited number of research works have documented the classification of PQDs through the implementation of ML models on FPGA such as the ones reported in \cite{FPGA-based_statistics,FPGA-based_CNN,LMS-LMF,LSSVM}. Consequently, the exploration of PQDs with FPGA-based ML implementations is ongoing and requires further development. In \cite{FPGA-based_statistics} the authors used statistical features such as the mean, variance, skewness and kurtosis from PQDs in conjunction with an artificial neural network implemented on FPGA to classify PQDs. This study also describes the implementation of specific hardware architectures to compute the statistical operations on FPGA. In \cite{FPGA-based_CNN} a signal processing technique called self-adaptive variational mode decomposition is integrated with deep CNN (DCNN) and another classifier called online-sequential random vector functional link network to classify PQDs on FPGA. In a study reported in \cite{LMS-LMF} neural networks based on the LMS and Least Mean Fourth control algorithms were deployed on FPGA for mitigation of harmonics and compensation of reactive power.
The authors in \cite{LSSVM} carried out the real-time detection and classification of evolving PQDs on an experimental setup where a signal processing technique called the energy preserving ensemble empirical mode decomposition in conjunction with a Least-Square SVM (LSSVM) were deployed on FPGA.

\section{Modeling Methodology\label{sec:model}}

In aircraft, a power monitoring system that can detect electrical faults and PQDs is necessary to ensure the safety and reliability of onboard systems. Aircraft power systems which operate under unique conditions such as high-frequency electrical signals and varying loads are susceptible to faults and PQDs like short and open circuits, voltage sags, transients or harmonic distortion. If undetected, these faults or disturbances can cause malfunction in critical systems such as navigation, communication or flight control that can potentially compromise both aircraft performance and safety. Real-time diagnostics are provided by the monitoring system which allows early detection of issues, corrective actions to be taken and preventive measures to be applied before faults escalate. Through continuous monitoring of PQ and identification of faults: operational integrity is maintained, the life of sensitive components is extended and compliance with aviation safety standards is ensured.

Integrating the detection of faults and PQDs into a power monitoring system in aircraft is critical to maintain the integrity and reliability of the aircraft’s electrical systems. Aircraft power systems responsible of feeding vital systems such as avionics, navigation, flight control, and communication, can face significant issues if PQ is disrupted or electrical faults occur. Any such disturbances, if not detected, can lead to system malfunctions, reduced performance, or safety risks.

By integrating fault and PQD detection, early identification of potential issues is enabled allowing corrective actions to be taken before escalation into severe problems. For example, voltage sags, harmonics, or transient disturbances can be detected to prevent damage to sensitive equipment that operates under strict voltage and frequency requirements. Faults, such as short circuits or overcurrent can be identified to minimize the risk of power distribution system damage and reduce the likelihood of complete system failure.

Additionally, real-time analysis of PQ across the aircraft is facilitated which helps to ensure that the power system remains within required operational specifications for safe flight. Valuable diagnostic information can be provided to enable predictive maintenance, to reduce downtime and to extend the lifespan of critical electrical components.

\subsection{Aircraft Power System Model}

\begin{figure}
\centerline{\includegraphics[width=\linewidth]{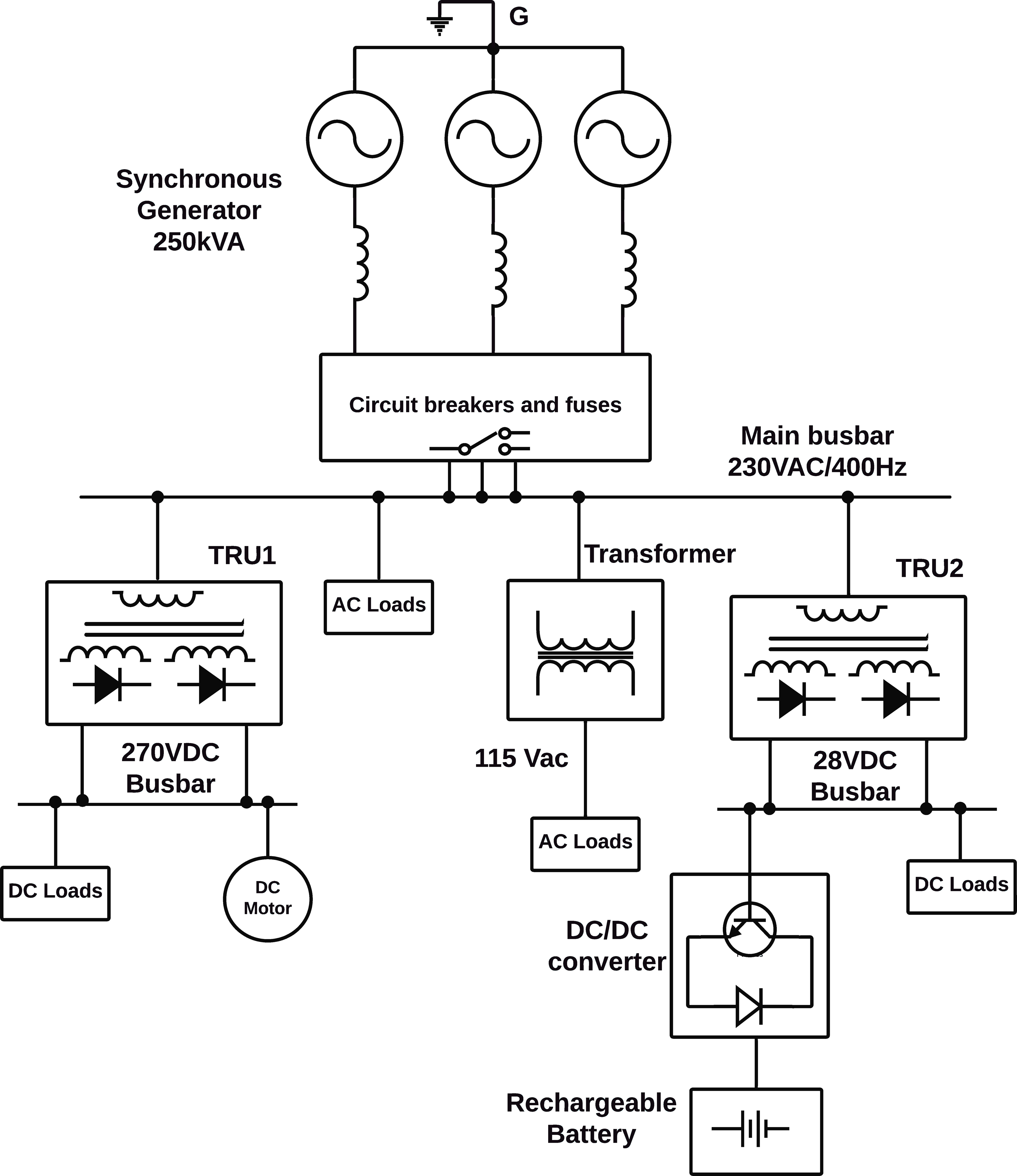}}
% \captionsetup{labelformat=empty}
\caption{Simulated power system model of the Boeing 787 aircraft.}
\label{PS}
\end{figure}

The power system model considered in this research is shown in \cref{PS} and draws inspiration from the electrical architecture of the Boeing 787 aircraft\cite{ref2}. The system incorporates a high-frequency 230V AC main busbar operating at 400Hz, alongside two secondary DC distribution lines delivering 28V and 270V, respectively. A 250~kVA, three-phase synchronous generator forms the primary energy source and interfaces with a centralized distribution unit. This unit is equipped with switching devices such as contactors and protective components including circuit breakers, which collectively manage the routing and safety of power delivery to various onboard electromechanical systems.

Power from the generator is fed into the 230V AC main busbar, which supplies a set of AC loads comprising both resistive elements and a motor. A transformer is integrated to step down the voltage for a 115V AC auxiliary busbar, enabling support for subsystems requiring lower AC voltage. Additionally, the system includes two transformer rectifier units (TRUs) (TRU1 and TRU2) that convert the main AC voltage into DC outputs at 28V and 270V. These DC buses serve different load types: the 28V line powers resistive loads and a DC/DC converter for regulated battery charging, while the 270V bus is used for both resistive components and a DC motor application.

Operational events such as switching actions or reconfigurations within the power distribution network lead to transient disturbances in both voltage and current signals. These fluctuations, often rich in spectral and temporal detail, are valuable for system diagnostics, as they reveal information about load types, switching dynamics, and operational conditions~\cite{ref3}. Furthermore, the occurrence of electrical faults, including short circuits and open circuit scenarios, generates distinctive transient profiles that can be exploited for fault detection and classification.

To capture and analyze these phenomena, a series of simulated fault scenarios were executed within the modeled power system. High-resolution signal recordings were obtained at a sampling rate of $f_s = 33.3$KHz to ensure detailed representation of the transient behaviors. 

\subsection{Simulated Operating Conditions}

\begin{figure*}[!htbp]
\centering

\includegraphics[width=\textwidth]{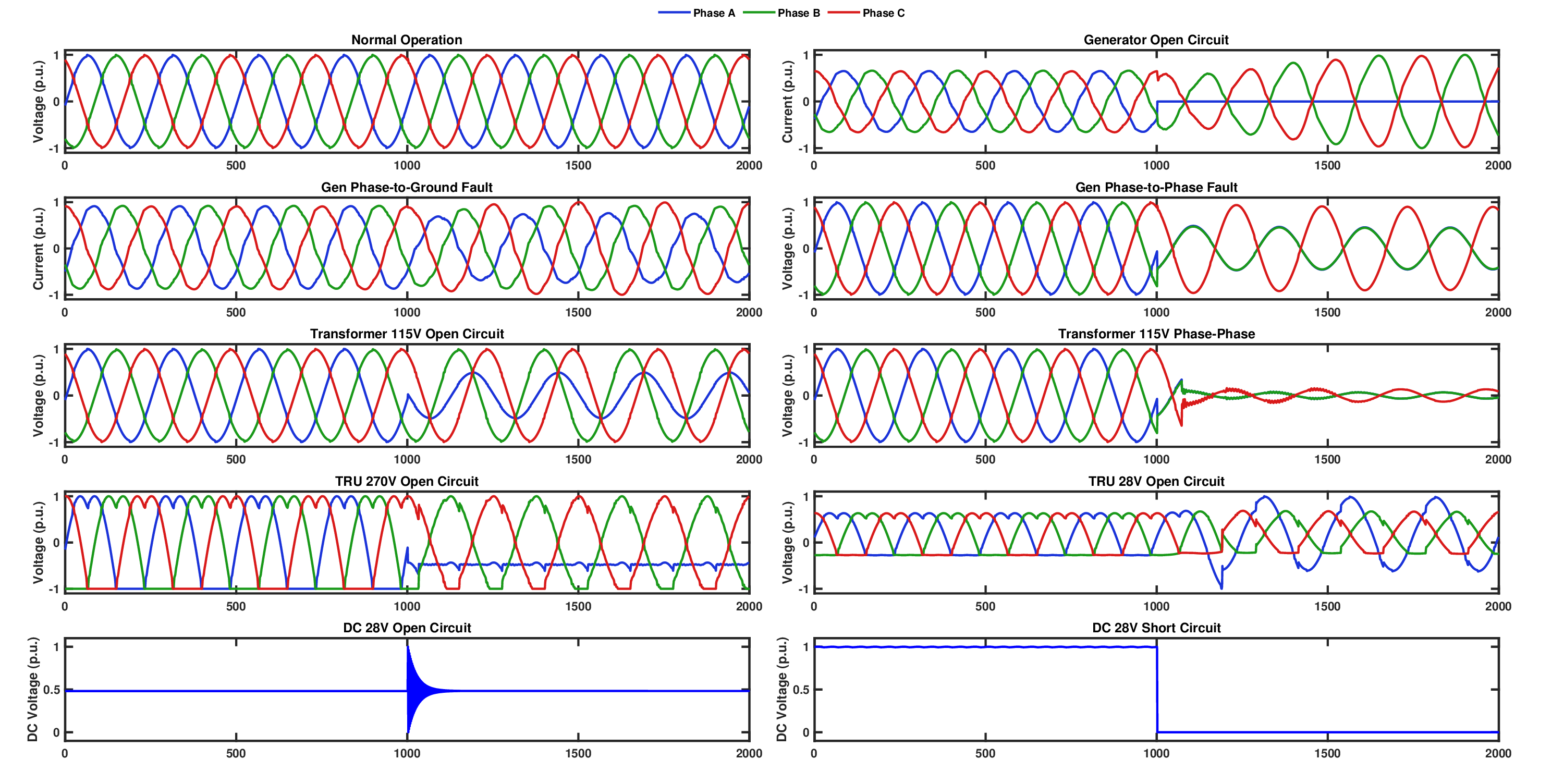}

\vspace{0.2cm}

\includegraphics[width=\textwidth]{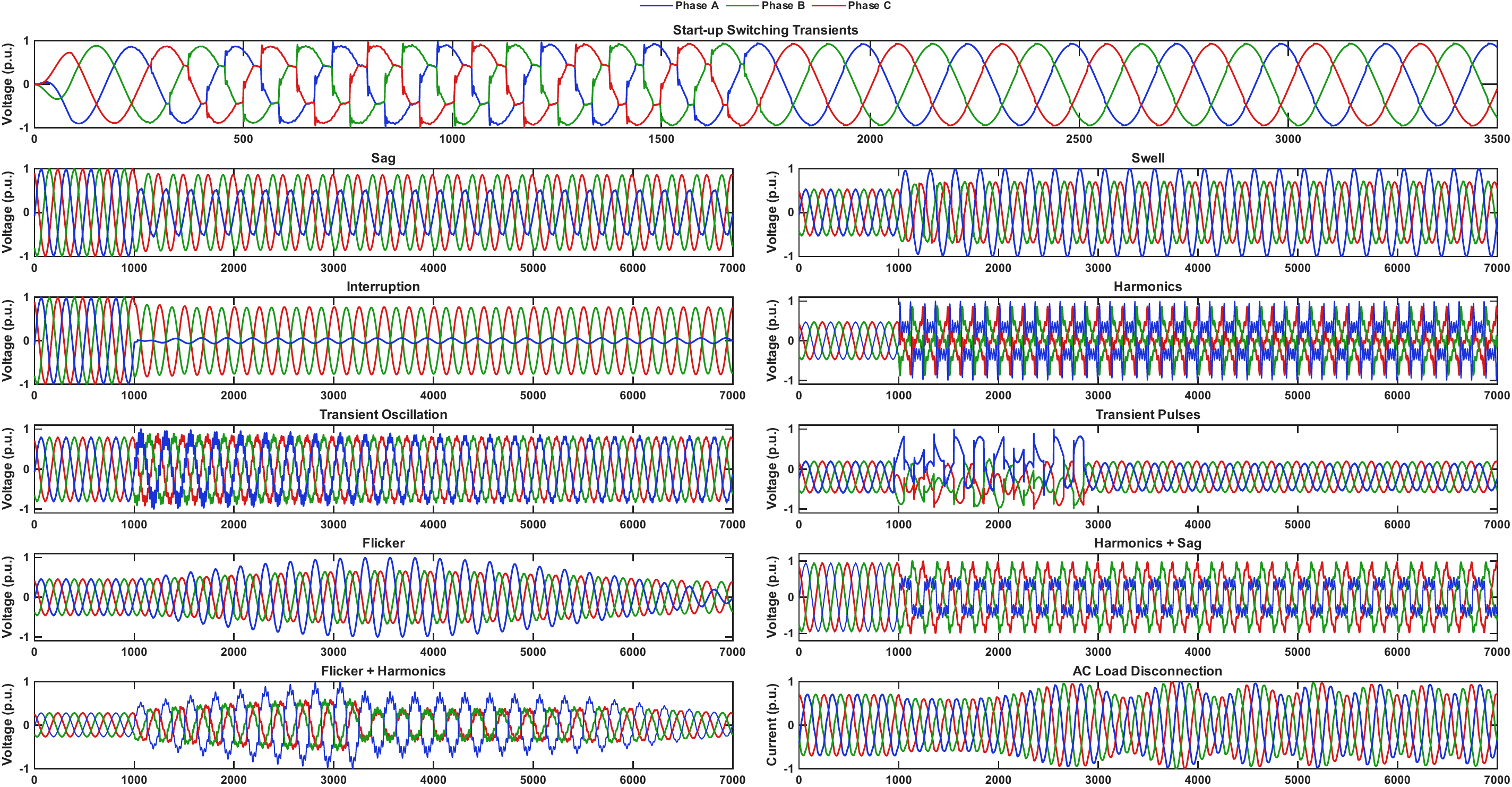}

\caption{Simulated fault events and PQDs in the Boeing 787 power system model under noise‑free conditions ($f_s = 33.3$kHz). Top: Fault events.  Bottom: Power quality disturbances. }
\label{fig:dist_faults}
\end{figure*}

The fault scenarios and PQDs simulated in the aircraft power system model are summarized in \cref{tab:classes}. Examples of the resulting signal data under noise‑free conditions are shown in \cref{fig:dist_faults}, where all faults and PQDs were simulated in phase A. The signals were captured using a multisensor approach. For each fault condition, the measurement point corresponds to the location indicated by the fault title in \cref{fig:dist_faults}. The transformer‑rectifier units (TRUs) exhibit natural harmonics that arise during the simulation of their corresponding fault conditions, whereas PQDs were simulated and measured at the main busbar. The simulated measurements use a multisensor formulation in which the signal modality is selected according to the physical observability of each event. Most PQDs and component-level fault cases are represented by voltage waveforms. However, the generator open-circuit and generator phase-to-ground fault cases, corresponding to Classes 13 and 14, are represented by current waveforms because these events produce more discriminative generator-current transients in the simulated aircraft power system depicted in \cref{PS}. The AC load disconnection case is also represented by a current waveform. This voltage/current measurement strategy is consistent with an onboard power-monitoring setting in which both voltage and current sensing channels are available. The model is inspired by the Boeing 787 electrical power system architecture and captures the dynamic behavior of aircraft power distribution, in contrast to conventional terrestrial power grids.

\begin{table}
\centering
\caption{Simulated PQDs and Fault Scenarios}
\label{tab:classes}
\begin{tabular}{|c|>{\centering\arraybackslash}p{6.4cm}|}
\hline
\textbf{Class} & \textbf{Description} \\
\hline
1 & Normal operation \\
2 & Generator start-up switching transients \\
3 & Voltage sag \\
4 & Voltage swell \\
5 & Voltage interruption \\
6 & Harmonics \\
7 & Flicker\\
8 & Transient oscillations \\
9 & Transient pulses \\
10 & Flicker + Harmonics \\
11 & Harmonics + Sag \\
12 & AC load disconnection \\
13 & Fault 1: Open circuit in the generator \\
14 & Fault 2: Phase-to-ground short circuit in the generator  \\
15 & Fault 3: Phase-to-phase short circuit in the generator  \\
16 & Fault 4: Open circuit in the 230/115~VAC transformer  \\
17 & Fault 5: Short circuit in the 230/115~VAC transformer \\
18 & Fault 6: Open circuit in the 270~V TRU \\
19 & Fault 7: Open circuit in the 28~V TRU \\
20 & Fault 8: Short circuit in the 28~V busbar \\
21 & Fault 9: Open circuit in the 28~V busbar \\
\hline
\end{tabular}
\end{table}
% These signals were generated using the model discussed in \cref{PS} and correspond to voltage measurements except for the AC load disconection which is a current waveform, which is inspired by the Boeing 787 electrical power system architecture and captures the dynamic behavior of aircraft power distribution, as opposed to conventional terrestrial power grids.

Differences between the simulated signals presented in this work and those commonly reported in the literature arise primarily from the underlying system model and signal generation strategy. Unlike many existing datasets, which are derived from conventional utility‑scale power grids operating at a nominal frequency of 60Hz, this work considers an aircraft power system model that operates at a higher fundamental frequency of 400Hz. This higher operating frequency influences the spectral content and transient behavior of both fault conditions and power quality disturbances, resulting in signal characteristics that differ from those commonly reported in grid‑based studies. This approach yields signal characteristics that differ from those in the literature but better reflect realistic operational variability. In addition, parameter diversity was intentionally introduced through domain randomization, with broad distributions applied to signal depth and duration, harmonic content, transient frequency and damping characteristics, noise level, and frequency deviation. The simulations further account for load switching transients and sampling jitter, all of which contribute to increased signal variability.
These modeling choices result in signal characteristics that more closely reflect the operational conditions encountered in aircraft electrical systems, while explaining observed discrepancies with grid‑based datasets commonly used in prior studies. The resulting signals therefore provide a realistic and challenging dataset for evaluating fault and PQD classification performance in aerospace‑relevant environments. While similar signals are extensively investigated in the context of conventional 50/60 Hz power grids, such as in \cite{paper_suggested_reviewer}, this research investigates their behavior in aircraft electrical systems which typically operate at 400 Hz. This higher operating frequency along with the compact and mission-critical nature of aircraft power systems introduces unique challenges not typically addressed in traditional power system studies. The simulated operating conditions and PQDs considered in this work include normal operation and generator start-up switching transients, as well as voltage sag, voltage swell, and voltage interruption. In addition, waveform distortions such as harmonics, flicker, transient oscillations, and transient pulses are modeled. Combined disturbances, including flicker with harmonics and harmonics with voltage sag, are also included to reflect realistic non-ideal operating scenarios. Furthermore, load-related events such as AC load disconnection are considered, along with a comprehensive set of fault conditions that affect key components of the aircraft electrical power system, including open- and short-circuit faults in the generator, transformers, transformer rectifier units (TRUs), and busbars. These scenarios aim to capture a wide range of operational, transient, and fault conditions representative of aerospace electrical power systems.

\subsection{Signal Processing--Based Data Augmentation}

 Data augmentation techniques based on signal processing methods such as rotation, scaling, frequency deviation, sampling jitter, and the addition of noise such as additive white Gaussian noise (AWGN) and color noise were applied to generate the majority of the dataset samples as described in \cite{Anempiricalsurveyofdataaugmentation}. In \cite{EnhancingTimeSeriesData}, the application of these three techniques for data augmentation led to improved model performance.
 A generated $N$-sample voltage waveform $\mathbf x \in \mathbb R ^ {N \times 1}$ is first rotated to obtain a second copy, where rotation for univariate time series is implemented using a negation in the literature, doubling the number of waveform examples (a positive $\mathbf x^+ \in \mathbb R ^ {N \times 1}$ and a negative copy, $\mathbf x^- \in \mathbb R ^ {N \times 1}$, collectively referred to using $\mathbf x$ this point onward). This approach acts as a form of regularization, similar to adding a penalty term to the loss function \cite{TrainingwithNoise}. Afterwards, to generalize the neural networks by preventing them from overfitting to the training data, scaling is applied to modify the global intensity of the time series by multiplying each sample by a random scalar value, and additive white Gaussian noise (AWGN) is added, as follows:
 
 \begin{equation}
      \mathbf{x}'= \alpha\mathbf{x}+\mathbf n
     % \mathbf{x}'=\diag\left(\mathbf s\right)\mathbf x + \mathbf n,
     \label{eq1}
 \end{equation}

 Where $\alpha$ is a random scalar value, $\mathcal{N}(\mu, \sigma^{2})$ denotes a Gaussian (normal) distribution with mean $\mu$ and variance $\sigma^{2}$, and $\mathbf{n} \in \mathbb{R}^{N \times 1} \sim \mathcal{N}(0, \sigma_{n}^{2})$ is the added AWGN (additive white Gaussian noise) vector.
For multivariate time-series data, a rotation (scaling) matrix can be applied as $\mathbf{x}' = \operatorname{diag}(\mathbf{s})\,\mathbf{x} + \mathbf{n}$, where $\mathbf{s} \in \mathbb{R}^{N \times 1} \sim \mathcal{N}(1, \sigma_s^{2})$ is the scaling matrix and $\operatorname{diag}(\cdot)$ denotes the diagonalization operator. Various values of $\sigma_s^{2}$ and $\sigma_n^{2}$ may be used to incorporate stronger or weaker impairments.

In addition to these signal processing--based techniques used to generate most of the dataset samples, the remaining samples were generated using GANs. GANs were introduced in 2014 in \cite{goodfellow2014generativeadversarialnetworks} and are DL models that generate new, realistic data by learning the underlying distribution of a given dataset. GANs were originally designed to generate image data but they can be adapted to generate time series data \cite{AssessingDeepGenerativeModelsonTimeSeriesNetworkData}. GAN-based data generation has also been investigated for PQD
classification as a way to address limited labeled data and improve
classifier training under data-scarce conditions \cite{jian2021ganpqd}.

\section{Experiments\label{sec:exp}}

\subsection{Dataset}

After generating the simulation data, data augmentation was applied using the previously described signal-processing techniques, including rotation, scaling, frequency warping, sampling jitter, and the addition of noise such as AWGN and color noise. Subsequently, GANs were trained to synthesize additional samples. Based on this procedure, two datasets were constructed. The first dataset was derived from the original time-series signals, each consisting of 1,000 samples, available on IEEE Data Port \cite{dataset}. The second dataset was generated by transforming the time‑series signals into their corresponding time–frequency representations using the STFT \cite{Continuous_Human_Action_Recognition_STFT_TAES}, With a window length of 64 samples, an overlap of 32 samples, and an FFT size equal to the window length, this yielded arrays of size 33 × 30.

\begin{table}
\centering
\caption{Per-class distribution of raw and augmented samples across 21 classes}
\label{tab:class_distribution}

\begin{tabular}{|c|c|c|c|c|}
\hline
\textbf{Class} 
& \textbf{Raw} 
& \textbf{SP-based} 
& \textbf{GAN-based} 
& \textbf{Total} \\
\hline

Each Class 
& 100 
& 3,000 
& 400 
& 3,500 \\
\hline

Dataset 
& 2,100
& 63,000
& 8,400 
& 73,500 \\
\hline

(Percentage) 
& (2.86\%) 
& (85.71\%) 
& (11.43\%) 
& (100\%) \\
\hline

\end{tabular}

\end{table}

Both datasets comprise 21 classes, as listed in \cref{tab:classes}, with 3,500 samples per class, resulting in a total of 73,500 samples. Each class includes raw simulation samples, signal-processing-based augmented samples, and GAN-generated samples, as summarized in \cref{tab:class_distribution}.Therefore, the data augmentation process was carried
out prior to splitting the dataset into training, validation, and testing sets. The dataset was then partitioned with proportions of 70\%, 10\%, and 20\%, corresponding to the training (51,450 samples), validation (7,350 samples), and testing (14,700 samples) subsets, respectively.

It is important to clarify the scope of this evaluation protocol. Since
signal-processing-based augmentation and GAN-based synthesis were
performed prior to the train/validation/test split, the reported test results
should be interpreted as a held-out-sample evaluation over stochastic
realizations drawn from the same simulated aircraft-power-system
data-generation process. Improper handling of related or transformed samples
across training and test partitions is a known source of overly optimistic
machine-learning performance estimates \cite{kapoor2023leakage}. Similar effects have been
reported when highly related samples or multiple derived instances from a
common acquisition source are allowed to cross validation partitions
\cite{tampu2022leakage,bussola2019tiles}. The augmented samples used
in this work are not deterministic duplicates of the raw signals; rather, they
are generated through randomized amplitude scaling, additive noise,
controlled time-warping, sampling jitter, low-frequency drift, localized
transient variation, and polarity inversion. Therefore, the test set does not
contain exact duplicate waveform copies observed during training. However,
this protocol is not equivalent to a parent-waveform-disjoint or independent
operating-campaign validation. Such validation, including grouped
parent-waveform splitting and experimental testbed evaluation under
independently generated operating conditions, is identified as future work.

\subsubsection{GAN-Based Data Generation}

% The proposed GAN architecture for data augmentation, depicted in \cref{GAN_Architecture}, was trained independently for each class in the dataset. 
% The proposed GAN architecture for data augmentation, shown in \cref{GAN_Architecture}, was trained independently for each class in the dataset. This class-specific training strategy enables the generator to capture the distinct temporal and spectral characteristics associated with each fault and disturbance type, thereby improving the fidelity of the synthesized signals. The generator network takes as input a latent noise vector and produces synthetic signals that aim to replicate the statistical properties of real samples within each class. The discriminator, in turn, is trained to distinguish between real and generated signals, encouraging the generator to produce increasingly realistic outputs through adversarial learning.

\begin{figure}
\centerline{\includegraphics[width=\linewidth]{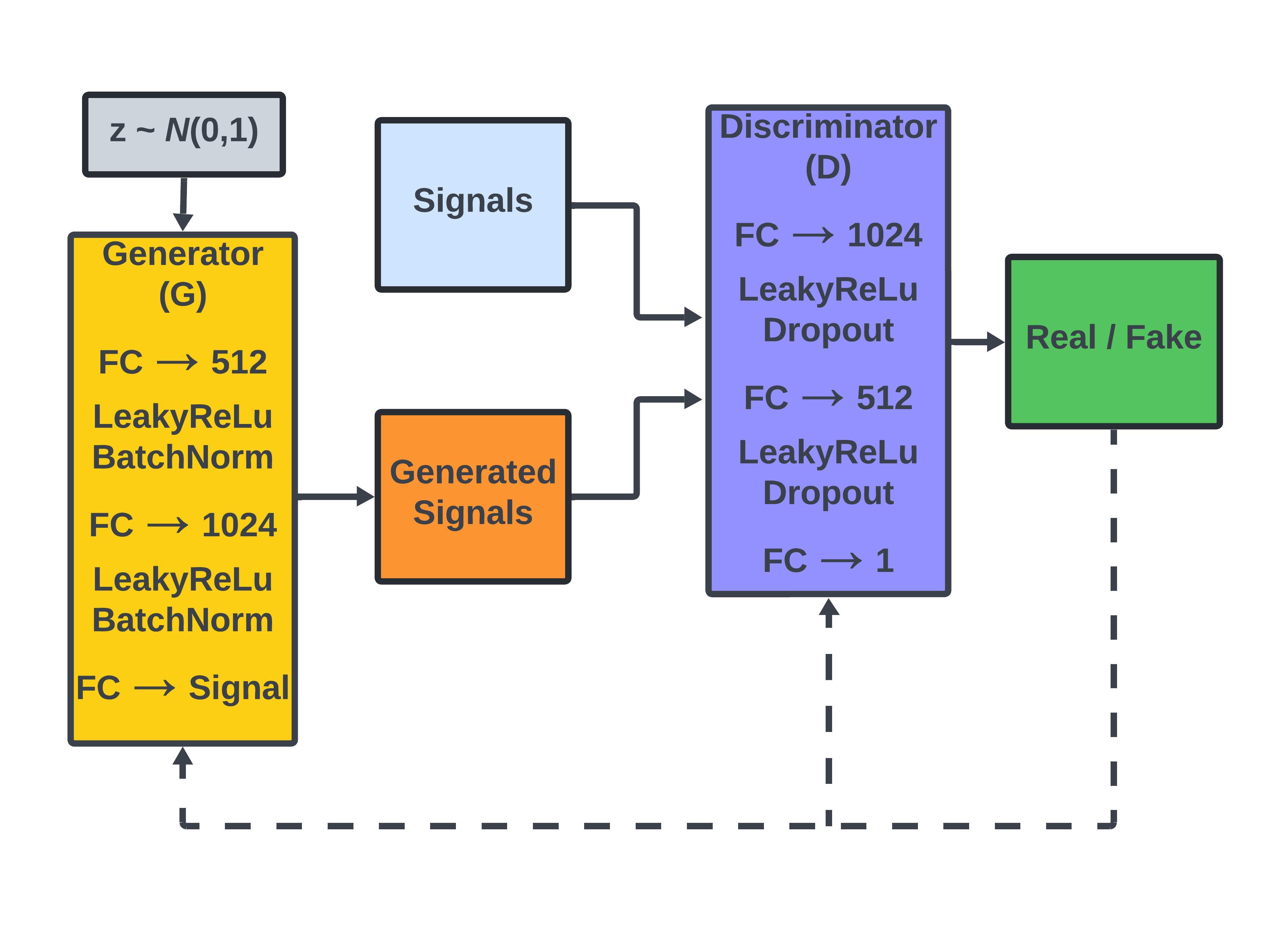}}
\caption{GAN architecture.}
\label{GAN_Architecture}
\end{figure}

The proposed GAN architecture for data augmentation, shown in \cref{GAN_Architecture}, was trained independently for each class in the dataset for 50 epochs. This class-specific training strategy enables the generator to capture the distinct temporal and spectral characteristics associated with each fault and disturbance type, which improves the fidelity of the synthesized signals. The generator network takes as input a latent noise vector and produces synthetic signals that aim to replicate the statistical properties of real samples within each class. The discriminator is trained to distinguish between real and generated signals, which promotes the generation of more realistic outputs through adversarial learning. Training is performed separately for each class to avoid mode mixing and to ensure that class-dependent features, such as harmonic content, transient behavior, and stochastic variations, are adequately preserved. This approach is particularly important for power system signals, where different signal classes exhibit significantly different temporal and spectral signatures.

The quality of the GAN-generated signals is quantitatively evaluated through complementary frequency-domain, time-domain, and feature-space analyses to ensure that the synthesized samples preserve the physical characteristics of aircraft electrical power systems while expanding dataset diversity.

\begin{figure}
    \centering
    \centerline{\includegraphics[width=\linewidth]{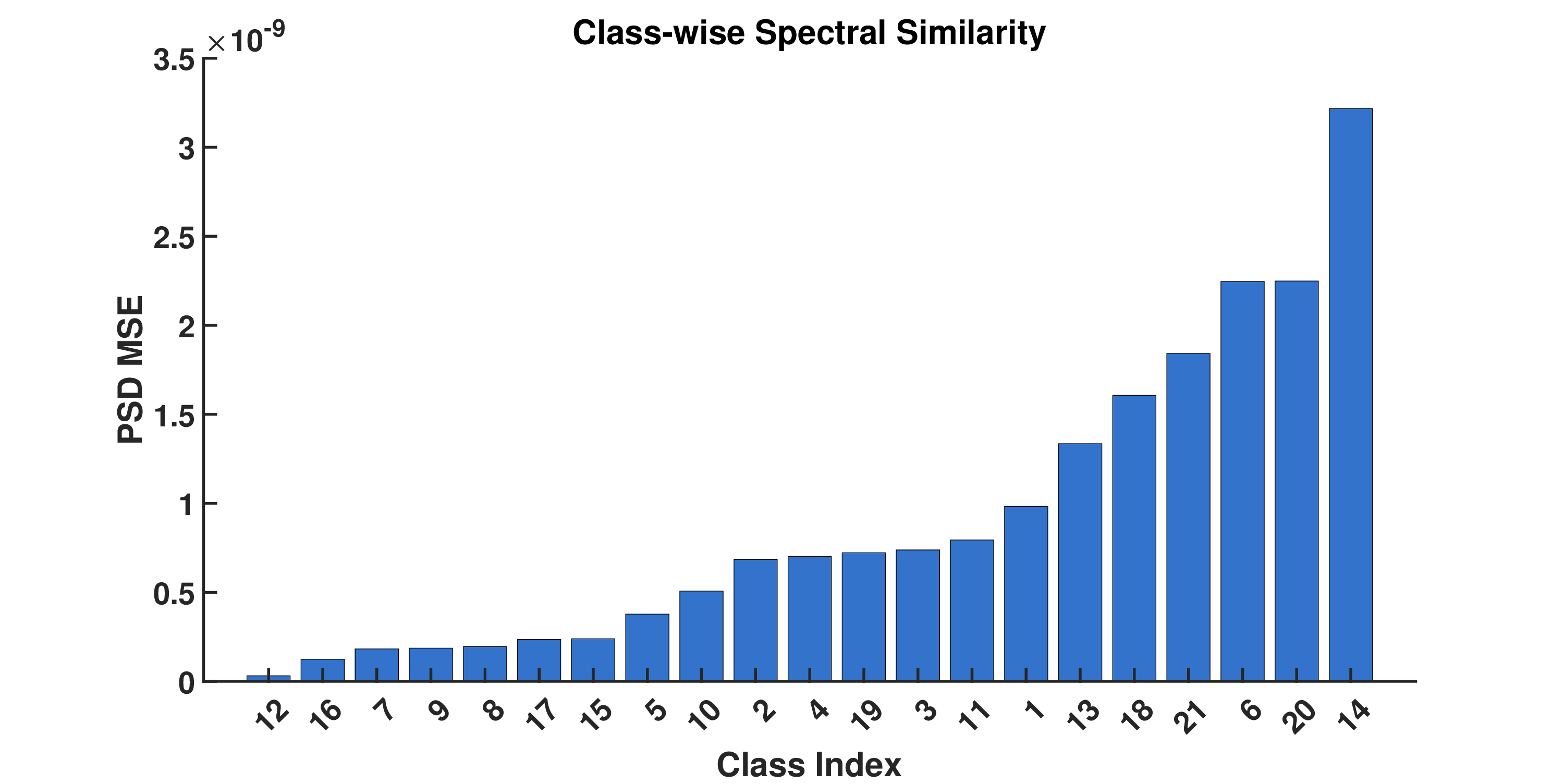}}
    \caption{Class-wise PSD similarity between real and generated signals.}
    \label{fig:psd_mse}
\end{figure}

First, the class-wise power spectral density (PSD) similarity between real and GAN-generated signals, shown in \cref{fig:psd_mse}, demonstrates very low spectral discrepancies across all 21 classes, with mean squared errors on the order of $10^{-9}$. The PSD similarity plot is presented in ascending order, which highlights the gradual variation in spectral agreement across fault and disturbance classes. This result confirms accurate preservation of dominant spectral features associated with aircraft power systems under 400Hz operation, including harmonic components, flicker-induced sidebands, and broadband transient energy. Consequently, the GAN maintains the spectral integrity required for reliable power quality and fault analysis.
% Slightly larger PSD deviations appear in disturbance classes dominated by transient and combined events, which inherently exhibit broader and more stochastic spectral content; however, these deviations remain within physically realistic bounds and do not indicate the presence of spurious frequency components. 

\begin{figure}
    \centering
    \centerline{\includegraphics[width=\linewidth]{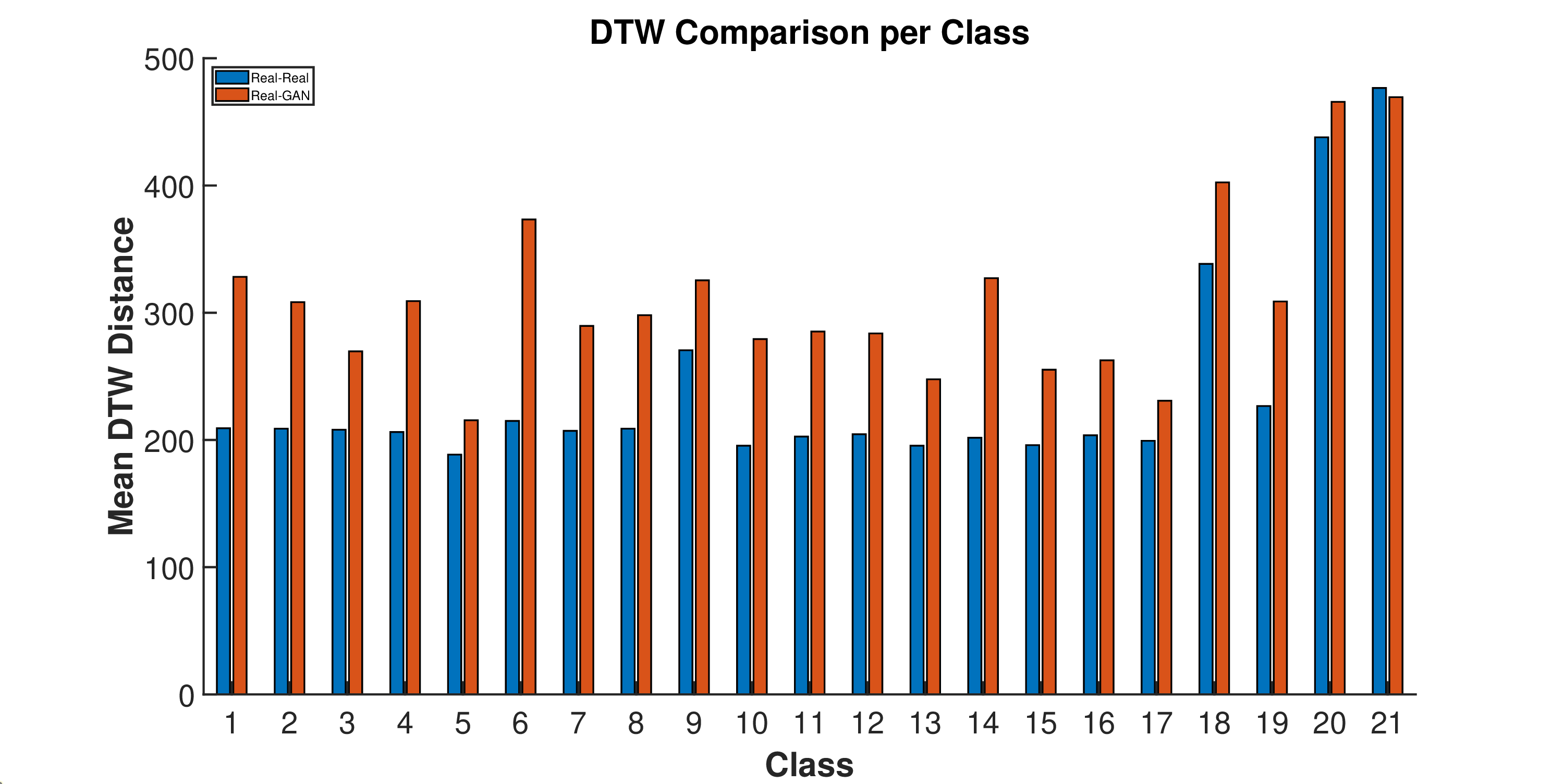}}
    \caption{Class-wise DTW Distribution.}
    \label{fig:dtw}
\end{figure}

Temporal consistency is further assessed through a class-wise dynamic time warping (DTW) analysis, presented in \cref{fig:dtw}. The results indicate that the mean DTW distances between real and GAN-generated signals closely follow the corresponding real-to-real baseline across all classes, with consistent relative ordering. Disturbance classes associated with abrupt waveform changes and severe transients, such as voltage interruptions and fault events, yield higher DTW values in both comparisons, whereas smoother disturbances produce lower values. A moderate increase in the real-to-GAN DTW distances appears, which reflects controlled and physically plausible temporal variation rather than waveform distortion. It is emphasized that the objective of GAN-based data augmentation is not the creation of identical replicas, but the synthesis of realistic intra-class variability that improves model generalization. The DTW results therefore confirm preservation of temporal morphology, event timing, and transient evolution, while enriching the dataset with meaningful diversity.

\begin{figure}
    \centering
    \centerline{\includegraphics[width=\linewidth]{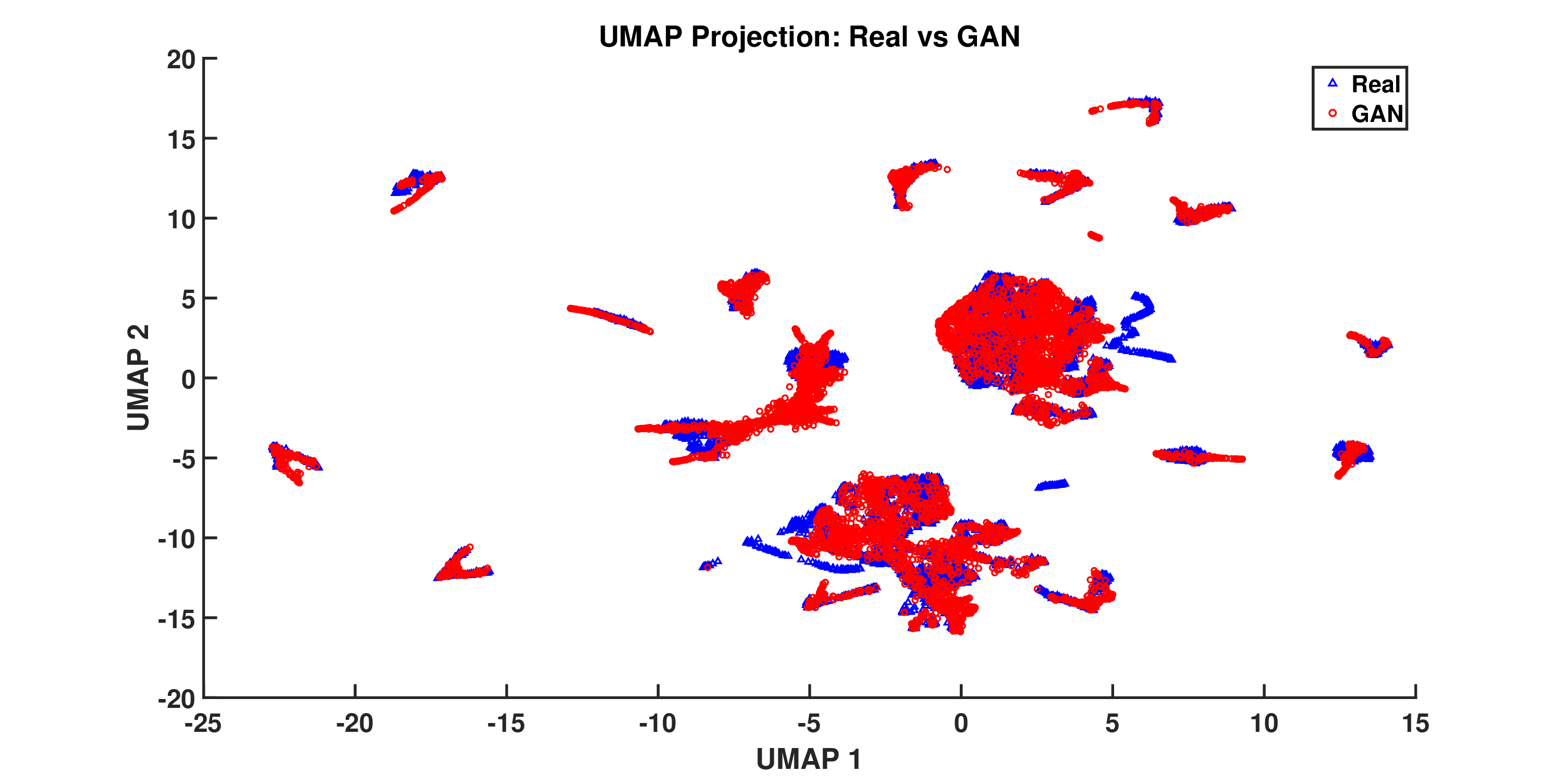}}
    \caption{UMAP Projection.}
    \label{fig:umap}
\end{figure}

Finally, the structural consistency of the augmented data is examined through an overall uniform manifold approximation and projection (UMAP) of real and GAN-generated samples, as shown in \cref{fig:umap}. The projection reveals substantial overlap between real and synthetic samples within the latent feature space, with well-defined clusters. GAN-generated samples occupy the same latent manifolds as the real data and extend coverage across these regions, without evidence of mode collapse or unintended mixing. This behavior indicates successful learning of the underlying feature distributions associated with normal operation, power quality disturbances, and fault conditions in aircraft electrical power systems. Collectively, these results demonstrate that GAN-based augmentation preserves spectral, temporal, and structural characteristics while expanding intra-distribution diversity, thereby providing a principled complement to signal-processing-based augmentation techniques in the construction of robust training datasets.

\subsection{Classification Model Development}%
Several DL architectures were evaluated to examine both classification performance and computational efficiency. The experiments included models such as 1D-CNNs, 2D-CNNs, LSTM networks, hybrid models combining CNNs with LSTM networks \cite{Deep_Learning_in_Aircraft_Design_TAES}, and standard deep architectures including ResNet, MobileNetV2, VGG16 and VGG19   \cite{Power_Line_Recognition_DL_TAES}.

To capture temporal or sequential patterns in the input data, 1D-CNN and LSTM architectures were employed. Variants combining 1D-CNN with LSTM layers were tested to incorporate both local feature representations and longer-term dependencies. For spatial feature extraction, 2D-CNN architectures were utilized, and hybrid 2D-CNN-LSTM models were introduced to integrate spatial and sequential feature learning in a unified framework.

Residual networks, including 1D-ResNet18 and standard 2D ResNet18 and ResNet50, were explored to assess the benefits of residual connections on performance and training stability. Lightweight architectures, such as MobileNetV2 and Compact ResNet, were evaluated for efficiency in terms of parameter count and inference time. Classical VGG networks (VGG16 and VGG19) served as benchmarks for deeper architectures with standard convolutional blocks, while a denoising 2D-CNN model was tested to study the impact of preprocessing combined with CNNs on classification performance.

For all models, key metrics; including parameter count, training and testing accuracy, precision, recall, F1 score, and inference time per sample; were recorded. This comprehensive evaluation enabled a detailed comparison of both accuracy and computational efficiency, highlighting trade-offs between model complexity and speed. To ensure fairness, all models were trained with identical hyperparameters for 20 epochs, a learning rate of $10^{-3}$, the cross-entropy loss function, and optimized with the Adam optimizer.

\begin{figure*}%
\centerline{\includegraphics[width=\linewidth]{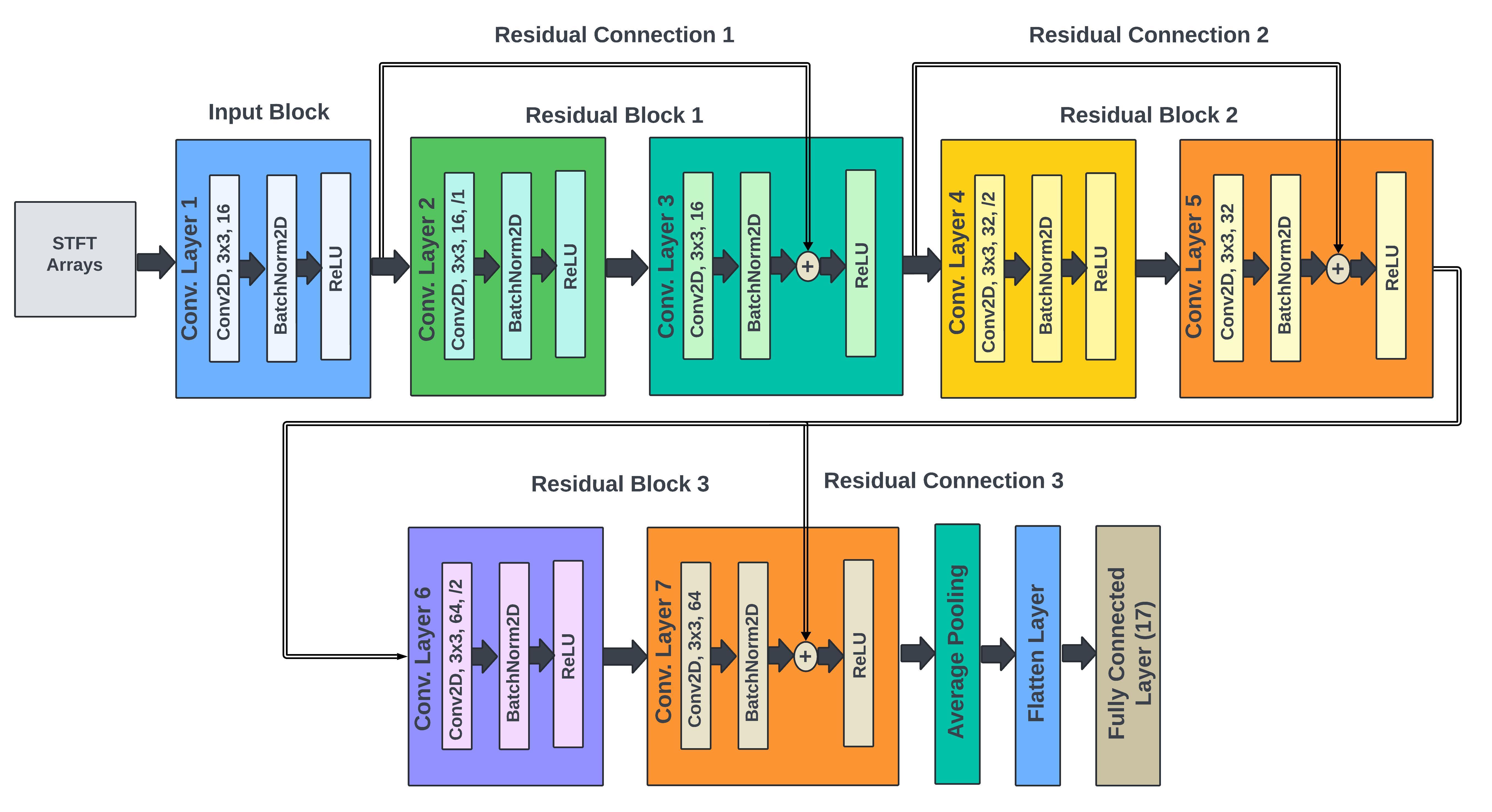}}
\caption{Compact ResNet architecture.}
\label{Compact_ResNet}
\end{figure*}

The proposed Compact ResNet model is based on the principles of residual learning originally introduced for image recognition tasks \cite{he2016deep} and operates on two‑dimensional input data. In this work, the input consists of STFT representations. The architecture employs residual connections that enable effective feature extraction while maintaining a low parameter count. The network comprises three residual layers with 16, 32, and 64 channels, respectively, followed by an adaptive average pooling stage and a fully connected output layer, as shown in \cref{Compact_ResNet}. In contrast to ResNet‑18, the Compact ResNet architecture targets lightweight deployment and contains exactly 175,685 parameters. This reduced model size supports efficient implementation on resource‑constrained FPGA platforms and aligns with the requirements of aerospace systems, where limitations on computational resources, memory capacity, and power consumption impose strict design constraints.

% The proposed compact ResNet model draws inspiration from residual learning for image recognition \cite{he2016deep} and incorporates residual connections to enable deep feature extraction with minimal parameters. The network consists of three residual layers with 16, 32, and 64 channels, respectively, followed by adaptive average pooling and a fully connected output layer as depicted in \cref{Compact_ResNet}. Unlike ResNet18, the Compact ResNet architecture was designed to be lightweight with only 175K parameters, making it suitable for aerospace applications where hardware resources are limited.

\subsection{Experimental Results}%

The comparative results of all evaluated models are summarized in \cref{tab:model_comparison}, including metrics such as the number of parameters, training and testing accuracy, precision, recall, F1 score, and inference time per sample. The table allows a detailed examination of the trade-offs between model complexity, computational efficiency, and classification performance, highlighting how lightweight architectures, hybrid models, and standard deep networks perform relative to one another. All the deep learning model development and training were conducted on Google Colab Pro servers using GPUs.%

\begin{table*}
\centering
\caption{Comparison of different models on classification performance and efficiency}
\resizebox{\textwidth}{!}{%
\begin{tabular}{cccccccc}
\hline
\textbf{Model} 
& \textbf{\# Params} 
& \makecell{\textbf{Test Acc.} \\ \textbf{(\%)}} 
& \makecell{\textbf{Precision} \\ \textbf{(\%)}} 
& \makecell{\textbf{Recall} \\ \textbf{(\%)}} 
& \makecell{\textbf{F1 Score} \\ \textbf{(\%)}} 
& \makecell{\textbf{Train Acc.} \\ \textbf{(\%)}} \\
% & \makecell{\textbf{Inf. Time} \\ \textbf{(ms/sample)}} \\
\hline
1D-CNN           & 8,338,709   & 97.29 & 97.32 & 97.29 & 97.30 & 98.41 \\ %& 0.02 \\
1D-CNN-LSTM      & 480,485     & 94.57 & 94.71 & 94.57 & 94.59 & 95.26  \\ %& 0.02 \\
1D-ResNet18      & 3,854,677   & 95.93 & 95.97 & 95.93 & 95.92 & 95.93  \\ %& 0.05 \\
2D-CNN           & 546,133     & 96.27 & 96.31 & 96.27 & 96.28 & 99.07  \\ %& 0.02 \\
2D-CNN-LSTM      & 755,925     & 95.68 & 95.73 & 95.68 & 95.69 & 96.94  \\ %& 0.02 \\
ResNet18         & 11,308,501  & 96.41 & 96.48 & 96.41 & 96.42 & 99.35  \\ %& 0.05 \\
ResNet50         & 23,544,789  & 95.88 & 95.95 & 95.88 & 95.90 & 99.28  \\ %& 0.11 \\
MobileNetV2      & 2,556,629   & 96.10 & 96.20 & 96.10 & 96.11 & 98.98  \\ %& 0.09 \\
VGG16            & 138,424,765 & 96.55 & 96.56 & 96.55 & 96.55 & 99.55  \\ %& 0.03 \\
Lightweight VGG19 & 20,302,165 & 96.48 & 96.51 & 96.48 & 96.48 & 99.49  \\ %& 0.04  \\
Denoising-Based 2D-CNN & 177,334  & 96.33 & 96.35 & 96.33 & 96.33 & 99.11  \\  %& 0.02  \\
\textbf{Compact ResNet (2D)} & 175,685 & 96.94  & 96.96 & 96.94 & 96.95 & 98.93  \\ %& 0.03  \\
\hline
\end{tabular}%
}
\label{tab:model_comparison}
\end{table*}

All evaluated models achieve test accuracies above 94\%, which confirms the effectiveness of the investigated architectures for the considered fault classification task. The highest test accuracy is obtained by the 1D-CNN, which reaches 97.29\%. However, this result requires 8.34 million parameters. The proposed Compact ResNet attains a test accuracy of 96.94\% with only 175,685 parameters, while the decrease in accuracy remains limited to 0.35 percentage points relative to the best-performing model. Such a small difference does not justify the substantial increase in computational cost associated with larger networks. Therefore, the Compact ResNet offers a more favorable balance between predictive performance and model efficiency.

\begin{figure}
\centerline{\includegraphics[width=\linewidth]{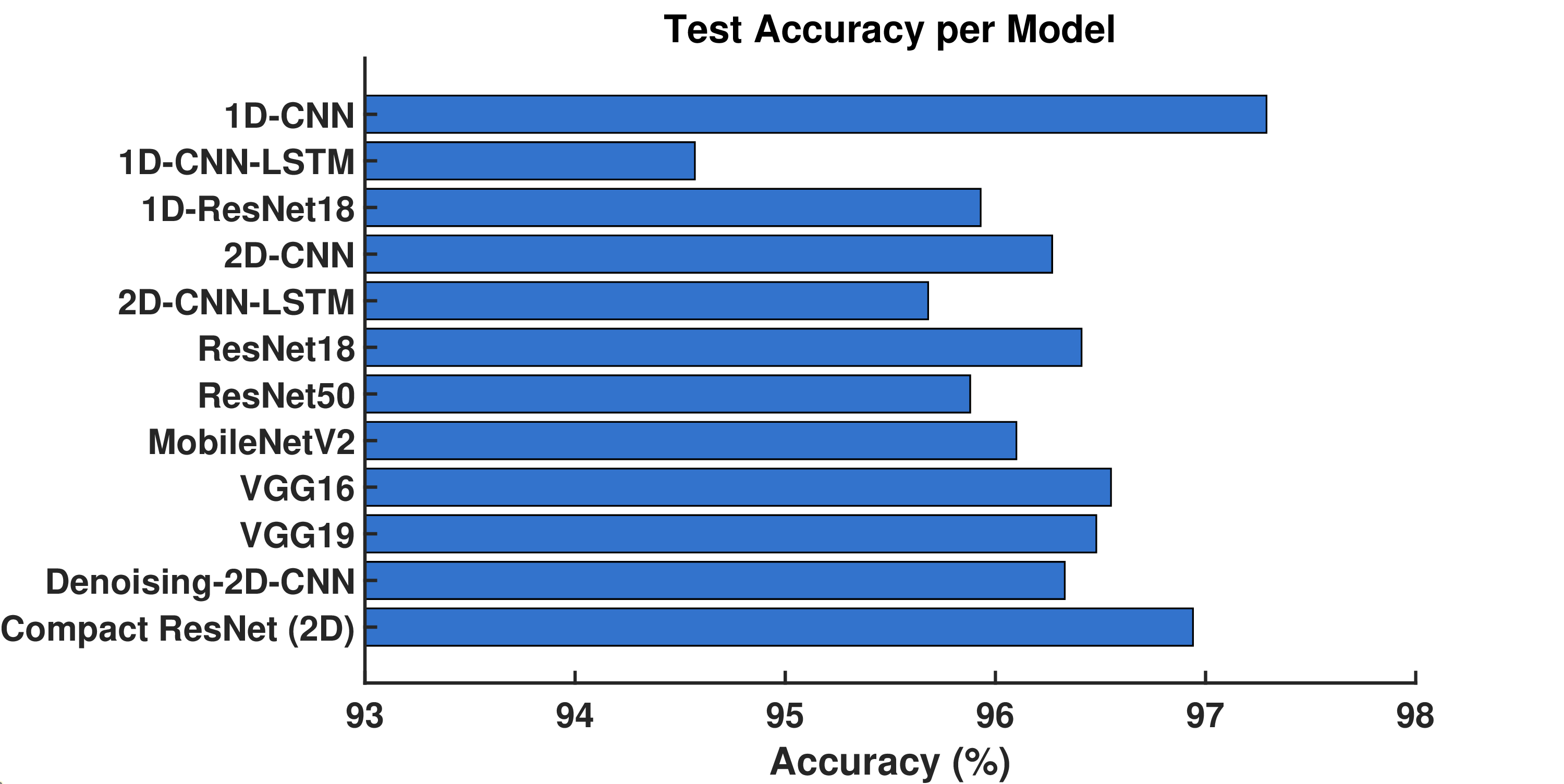}}
% \captionsetup{labelformat=empty}
\caption{Test set accuracy per model.}
\label{Test_acc}
\end{figure}

As illustrated in \cref{Test_acc}, the performance gap among the strongest models remains limited. VGG16 achieves a test accuracy of 96.55\%, ResNet18 reaches 96.41\%, Lightweight VGG19 attains 96.48\%, and the denoising-based 2D-CNN records 96.33\%. The Compact ResNet exceeds the performance of all these architectures and remains second only to the 1D-CNN. Under these conditions, model selection cannot rely solely on accuracy, and efficiency becomes a decisive factor. The Compact ResNet combines high predictive capability with the smallest parameter count among all evaluated models. This result indicates an efficient use of representational capacity. In contrast, the substantial increase in parameters observed in architectures such as ResNet50 and VGG16 does not produce proportional gains in accuracy. 

\begin{figure}
\centerline{\includegraphics[width=\linewidth]{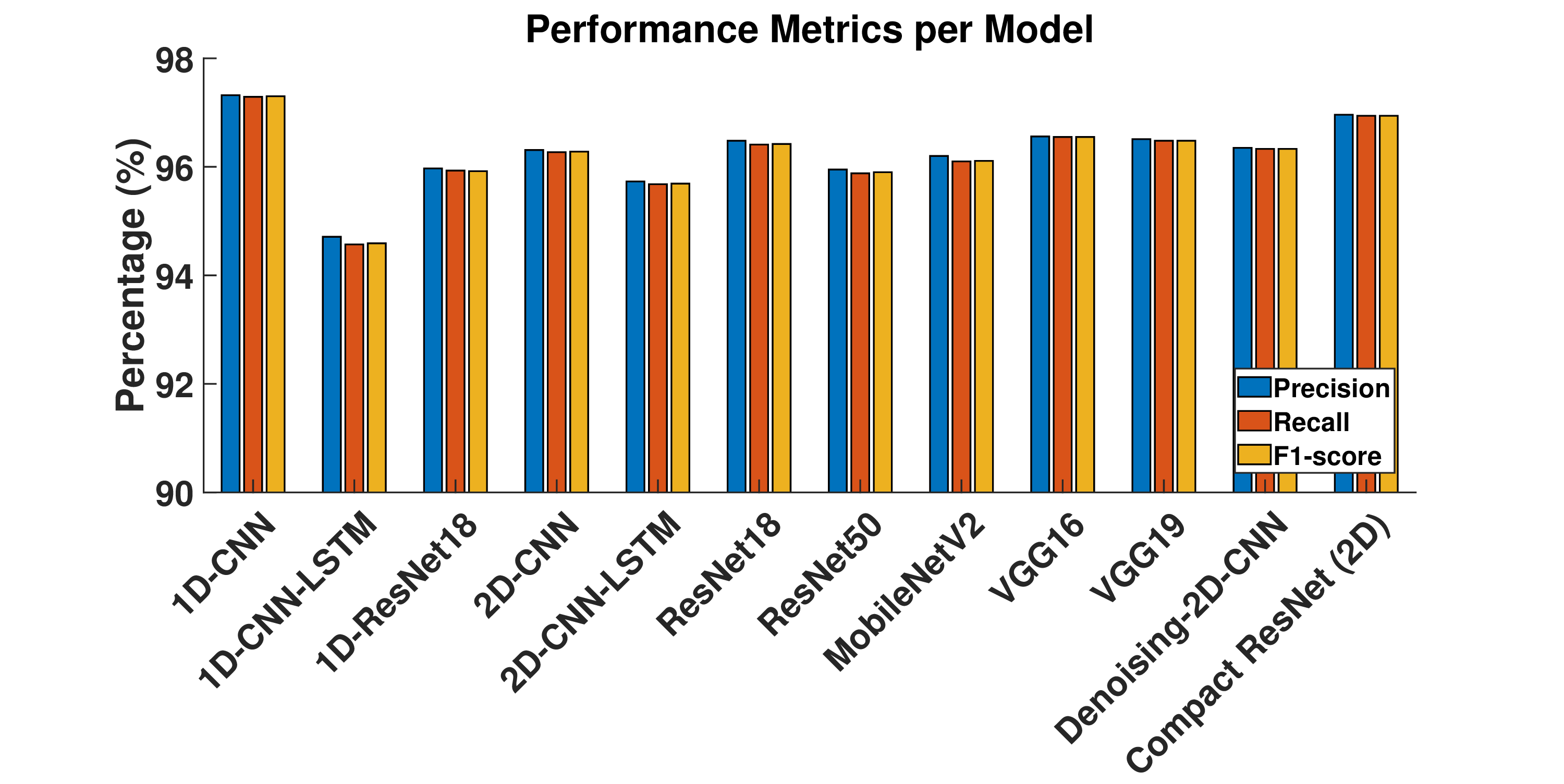}}
% \captionsetup{labelformat=empty}
\caption{Classification metrics.}
\label{Metrics}
\end{figure}

The classification metrics presented in \cref{Metrics} further support this observation. The Compact ResNet achieves precision, recall, and F1 score values of 96.96\%, 96.94\%, and 96.95\%, respectively. The close agreement among these metrics indicates balanced classification behavior across classes. The 1D-CNN exhibits the highest values, with 97.32\% precision, 97.29\% recall, and 97.30\% F1 score, but at the expense of a considerably larger model. In contrast, the lowest performance appears in the 1D-CNN-LSTM architecture, which records 94.71\% precision, 94.57\% recall, and 94.59\% F1 score despite the use of nearly half a million parameters. The 1D-ResNet18 and 2D-CNN-LSTM models also yield lower classification metrics than the proposed architecture. These results demonstrate that increased architectural complexity does not necessarily translate into improved predictive performance.

\begin{figure}
\centerline{\includegraphics[width=\linewidth]{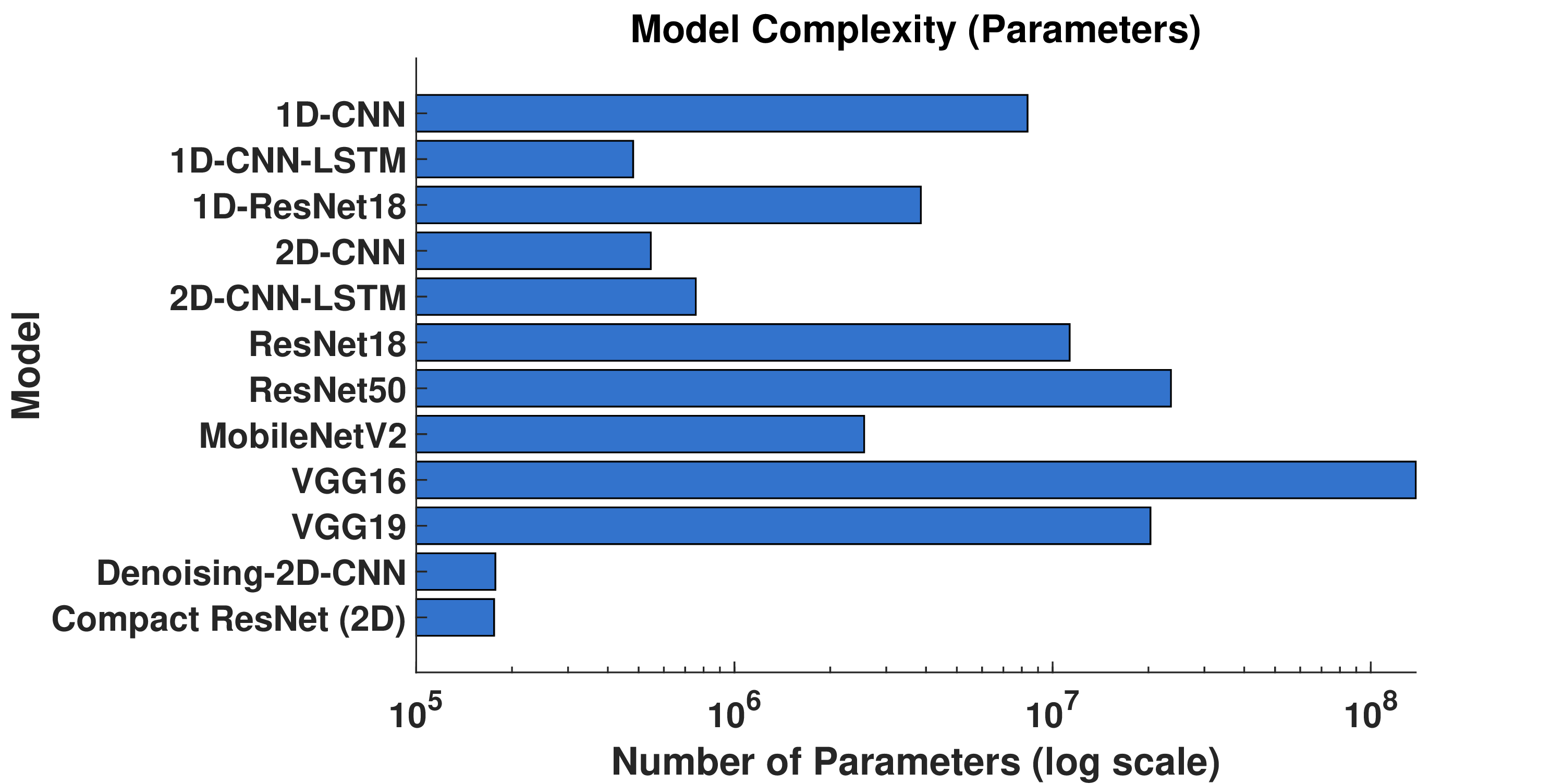}}
% \captionsetup{labelformat=empty}
\caption{Parameter count per model.}
\label{N_Parameters}
\end{figure}

Model complexity is further examined in \cref{N_Parameters}. The Compact ResNet constitutes the smallest architecture among all evaluated models, with only 175,685 trainable parameters. The denoising-based 2D-CNN represents the closest competitor in terms of compactness, with 177,334 parameters, but achieves a lower test accuracy of 96.33\%. Conversely, VGG16 requires approximately 138 million parameters and ResNet50 employs more than 23 million parameters, yet neither architecture provides meaningful performance improvements relative to the proposed model. This disparity highlights the inefficiency of excessively large networks for the considered application.

\begin{figure}
\centerline{\includegraphics[width=\linewidth]{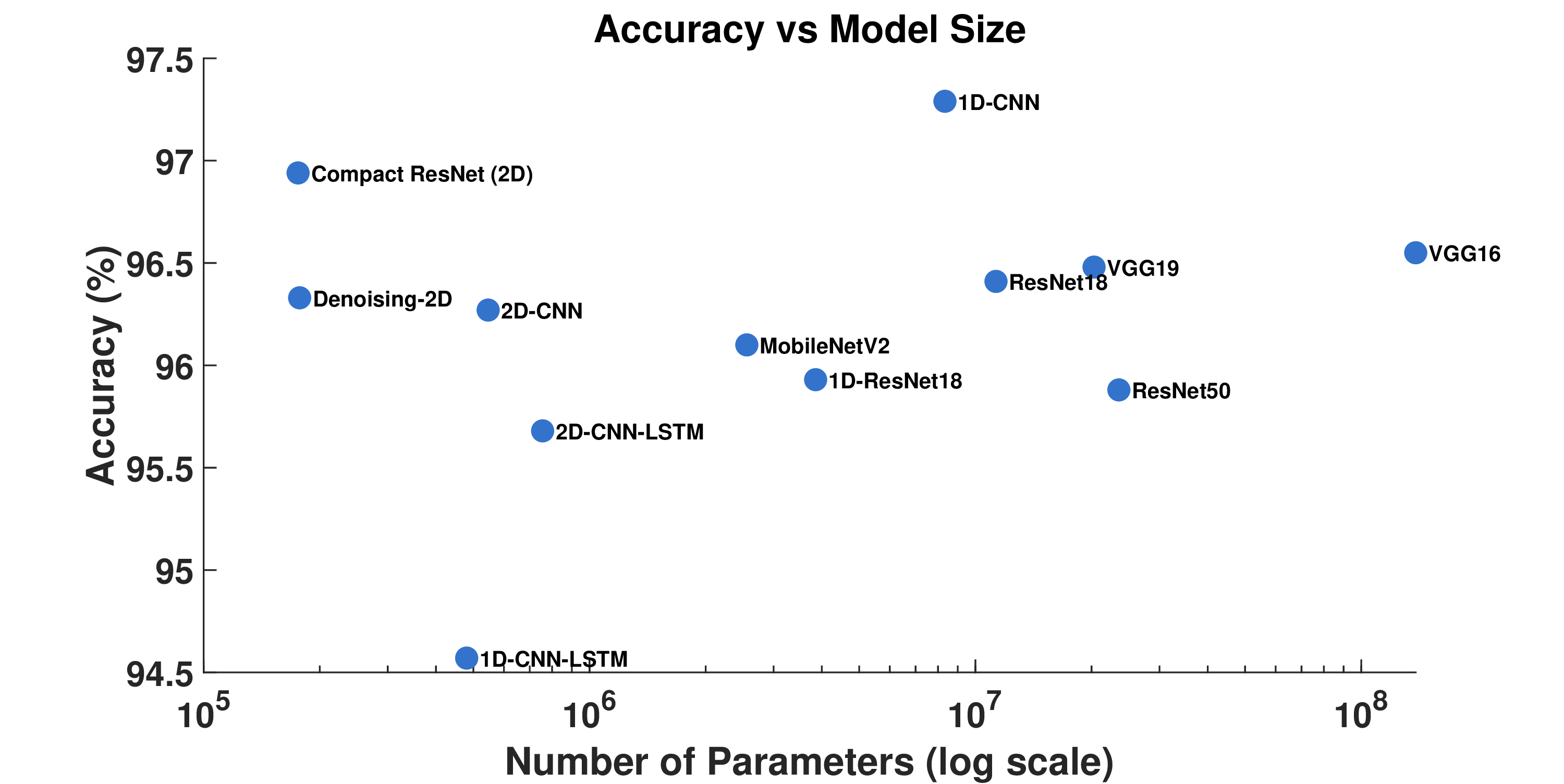}}
% \captionsetup{labelformat=empty}
\caption{Test accuracy vs. model size.}
\label{scatterplot_parameters}
\end{figure}

The relationship between test accuracy and model size is illustrated in \cref{scatterplot_parameters}. The Compact ResNet occupies the most desirable region of the accuracy--complexity space, where high predictive performance coincides with minimal parameter requirements. Larger architectures cluster in regions associated with substantially higher complexity without clear accuracy advantages. This distribution provides strong evidence of the superior efficiency achieved by the proposed architecture.

These findings have direct implications for deployment in aircraft systems, particularly on FPGA platforms. Architectures such as the 1D-CNN, ResNet18, ResNet50, Lightweight VGG19, and VGG16 require substantially larger memory resources because of their high parameter counts. Such requirements increase logic utilization, power consumption, and inference latency, all of which constitute critical constraints in FPGA-based real-time applications. Although the denoising-based 2D-CNN remains relatively compact, it still employs more parameters than the Compact ResNet and delivers lower classification performance. In contrast, the Compact ResNet preserves near state-of-the-art accuracy while maintaining balanced classification metrics and minimal computational cost. Its compact structure facilitates efficient allocation of FPGA resources, including DSP slices, LUTs, and BRAM, thereby enabling low-latency and energy-efficient inference suitable for real-time aircraft system deployment.

\subsection{Evaluation of Compact ResNet: Learning Curves and Confusion Matrix}

\begin{figure}
\centerline{\includegraphics[width=\linewidth]{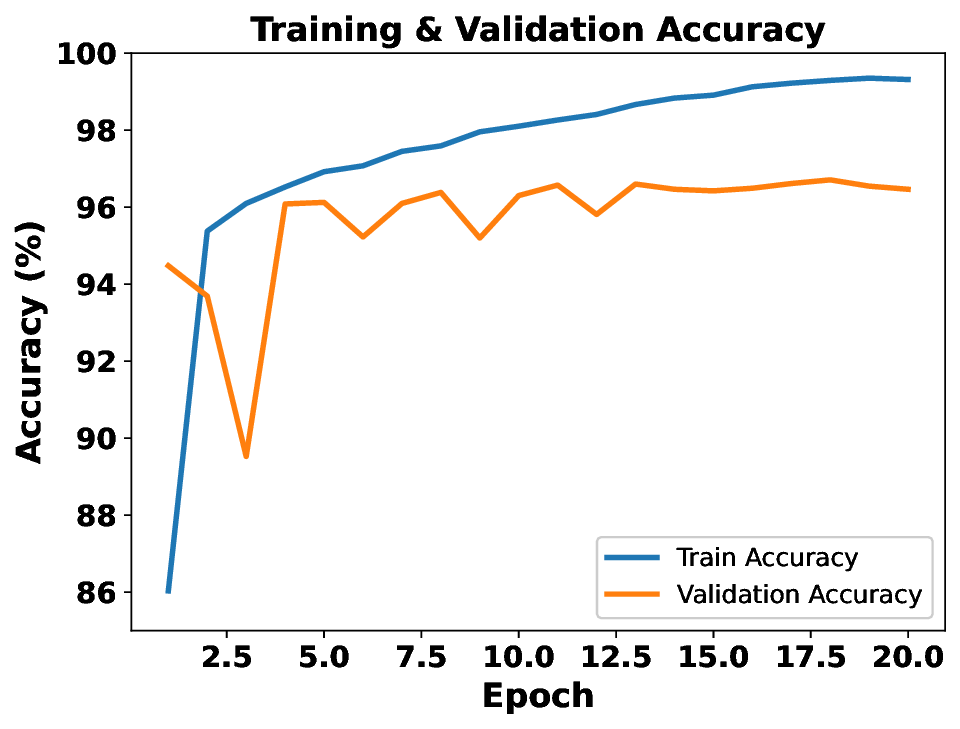}}
% \captionsetup{labelformat=empty}
\caption{Learning curves of the compact ResNet model.}
\label{learning_curves}
\end{figure}

\cref{learning_curves} presents the training and validation accuracy of the Compact ResNet model across 20 epochs. The model reaches rapid convergence during the early training phase, with training accuracy exceeding 90\% by the fifth epoch. This early performance indicates effective feature extraction and supports the efficiency of the proposed architectural design.

Validation accuracy follows a trajectory similar to that of the training curve after brief initial fluctuations and stabilizes between 95\% and 96\% from approximately epoch 13 onward. The consistently small gap between training and validation accuracy reflects strong generalization behavior and indicates a low risk of overfitting. Although slight variability appears in the validation results, overall accuracy remains stable across successive epochs.
This convergence pattern confirms stability in the optimization process and supports the suitability of the selected hyperparameters. Overall, these results validate Compact ResNet as a lightweight yet high‑performance architecture that achieves accurate classification while maintaining robust generalization, which supports suitability for deployment in resource‑constrained and real‑time environments.

\begin{figure}[!t]
    \centering
    \centerline{\includegraphics[width=\linewidth]{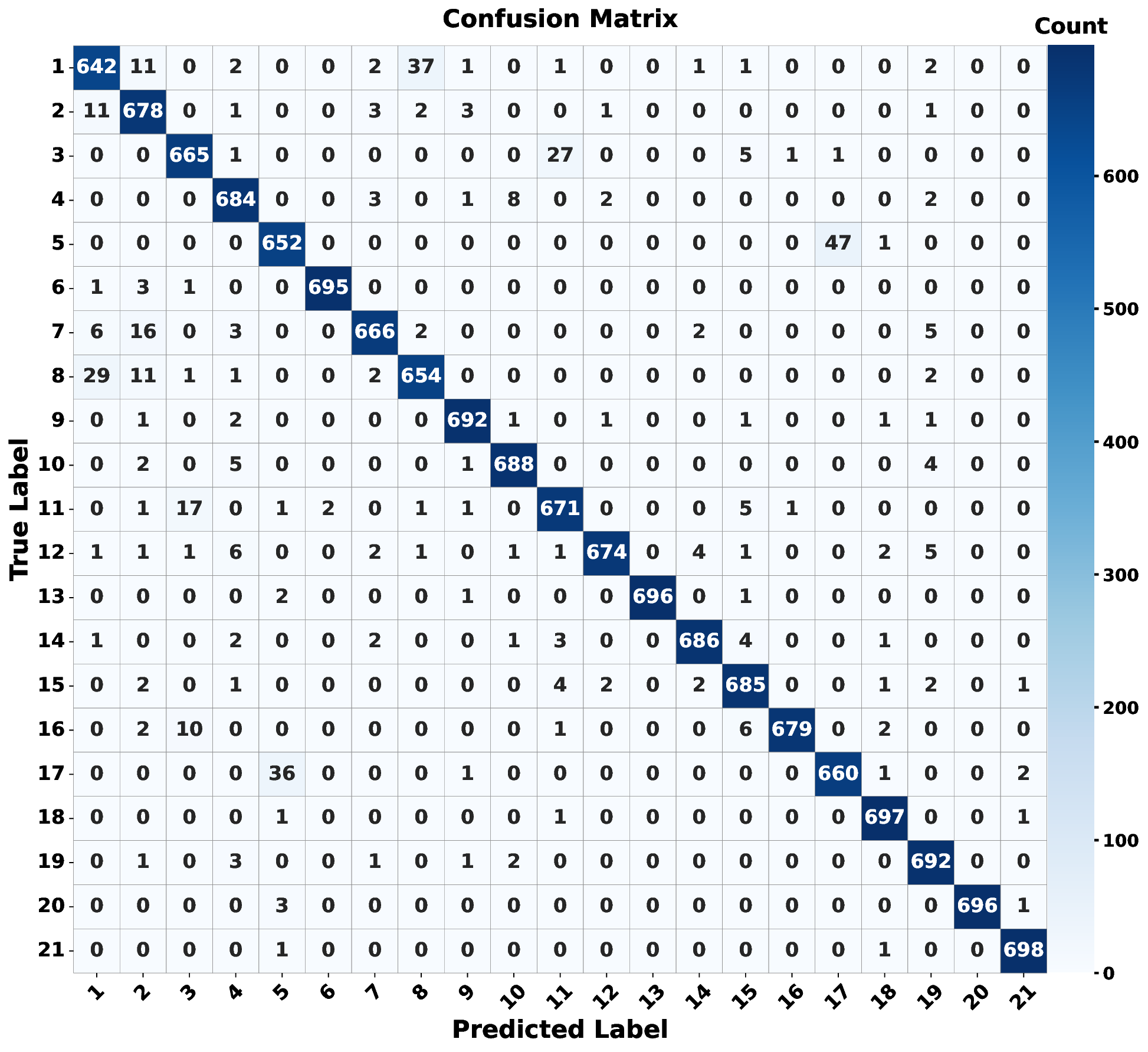}}
    \caption{Confusion matrix of the compact ResNet.}
    \label{confusion_matrix}
\end{figure}

The confusion matrix presented in \cref{confusion_matrix} provides a detailed assessment of the classification performance achieved by the proposed Compact ResNet architecture across all 21 classes. Strong concentration along the main diagonal is observed, where the majority of test samples are assigned to their correct class. For most classes, more than 650 of the 700 available test samples are correctly classified. This result confirms that discriminative class-specific features have been successfully extracted and utilized by the proposed model.

Only limited dispersion is observed within the off-diagonal elements of the confusion matrix. Classification errors are distributed across a small number of neighboring classes, while no dominant error pattern is evident throughout the complete label space. Since an equal number of test samples is allocated to each class, an unbiased evaluation of class-wise performance is provided. Furthermore, no class is associated with a disproportionately large contribution to the overall error distribution. Consequently, balanced classification behavior is achieved across the full set of operating conditions.

% The confusion matrix shown in \cref{confusion_matrix} demonstrates strong multi‑class classification performance across all 21 classes. A pronounced dominance of diagonal elements confirms that the majority of samples receive correct predictions. For most classes, diagonal counts exceed 650 samples, which reflects effective discrimination among class‑specific features. Several classes, including Classes 9, 18, 20, and 21, exhibit near‑perfect diagonal concentration with minimal off‑diagonal dispersion. These results align closely with the high recall values reported in Table \cref{tab:per_class_metrics_fnr_percentage} and validate the consistency of model predictions across a wide range of operating conditions. Each class in the test set contains 700 samples, which ensures a balanced basis for interpreting these results.

% From a global perspective, misclassifications remain limited and localized rather than widespread. Off‑diagonal entries primarily occur between operationally related or structurally adjacent classes, which is consistent with realistic fault‑classification behavior in avionics systems. No single class dominates the error distribution, which indicates balanced learning across the full label set. This balance remains essential in safety‑critical environments, where uneven class performance may lead to unacceptable operational risk.

Additional evidence of the robustness of the proposed approach is provided by the per-class metrics reported in \cref{tab:per_class_metrics_fnr_percentage}. Precision values above 92\% are obtained for all classes, while recall values exceeding 91\% are achieved throughout the entire dataset. For the majority of classes, recall values greater than 95\% are obtained. Similarly, consistently high F1-scores are reported across all classes, which confirms that a favorable balance between precision and recall has been maintained. Therefore, accurate fault identification is achieved without a corresponding increase in false alarm rates.

\begin{table}
\centering
\caption{Per-Class Precision, Recall, F1-Score, and False Negative Rate (FNR)}
\label{tab:per_class_metrics_fnr_percentage}
\begin{tabular}{c c c c c}
\hline
\textbf{Class} & \makecell{\textbf{Precision}\\\textbf{(\%)}} &
\makecell{\textbf{Recall}\\\textbf{(\%)}} &
\makecell{\textbf{F1-Score}\\\textbf{(\%)}} &
\makecell{\textbf{FNR}\\\textbf{(\%)}} \\
\hline
1  & 92.91 & 91.71 & 92.31 & 8.29 \\
2  & 93.00 & 96.86 & 94.89 & 3.14 \\
3  & 95.68 & 95.00 & 95.34 & 5.00 \\
4  & 96.20 & 97.71 & 96.95 & 2.29 \\
5  & 93.68 & 93.14 & 93.41 & 6.86 \\
6  & 99.71 & 99.29 & 99.50 & 0.71 \\
7  & 97.80 & 95.14 & 96.45 & 4.86 \\
8  & 93.83 & 93.43 & 93.63 & 6.57 \\
9  & 98.58 & 98.86 & 98.72 & 1.14 \\
10 & 98.15 & 98.29 & 98.22 & 1.71 \\
11 & 94.64 & 95.86 & 95.24 & 4.14 \\
12 & 99.12 & 96.29 & 97.68 & 3.71 \\
13 & 100.00 & 99.43 & 99.71 & 0.57 \\
14 & 98.71 & 98.00 & 98.35 & 2.00 \\
15 & 96.61 & 97.86 & 97.23 & 2.14 \\
16 & 99.71 & 97.00 & 98.33 & 3.00 \\
17 & 93.22 & 94.29 & 93.75 & 5.71 \\
18 & 98.59 & 99.57 & 99.08 & 0.43 \\
19 & 96.65 & 98.86 & 97.74 & 1.14 \\
20 & 100.00 & 99.43 & 99.71 & 0.57 \\
21 & 99.29 & 99.71 & 99.50 & 0.29 \\
\hline
\end{tabular}
\end{table}

Particular attention is given to the false negative rate (FNR), since missed detections may have significant consequences in safety-critical aerospace monitoring. Across all 21 classes, the FNR ranges from 0.29\% to 8.29\%. Therefore, the results indicate strong simulated-test performance, but further reduction of worst-class FNR under independent experimental conditions remains an important direction for future work.

% The performance obtained for the safety-critical classes, namely Classes 13 through 17 and Classes 20 and 21, is of particular importance. Recall values above 94\% are achieved for all classes within this subset, while recall values above 97\% are obtained for several classes. Only limited dispersion is observed within the corresponding rows of the confusion matrix, and the majority of samples remain concentrated along the principal diagonal. As a result, reliable identification of safety-relevant operating states is achieved, which is a critical requirement for practical avionics fault-diagnosis systems.

Another notable characteristic of the proposed model is the uniformity of performance across the complete class set. Significant variation between classes is not observed. Instead, consistently high precision, recall, and F1-score values are maintained throughout the dataset. Such behavior indicates that robust feature representations have been learned for a wide range of fault conditions and operational states. Consequently, stable performance may be expected when the proposed architecture is deployed under realistic operating scenarios.

Overall, the confusion matrix and the per-class evaluation metrics provide strong evidence of the effectiveness of the proposed Compact ResNet model. High recall values, low false negative rates, strong diagonal dominance, and consistently high F1-scores are achieved across all 21 classes. Collectively, these results support the suitability of the proposed architecture for avionics fault diagnosis and health-monitoring applications, where reliable and balanced fault classification is required.

\subsection{Evaluation of Model Robustness to Noisy Inputs}

Aerospace sensor measurements inherently include noise due to electromagnetic interference, environmental effects, and hardware limitations. To assess robustness under such conditions, this study evaluates classification performance under additive white Gaussian noise (AWGN) applied to the test dataset at multiple signal-to-noise ratio (SNR) levels. This experiment validates model reliability when exposed to degraded signal quality typical of real-world operating environments.

% The effect of additive noise on model performance was assessed across SNR levels from 0~dB to 30~dB, with results summarized in \cref{model_snr}. Under near-clean conditions at 30~dB SNR, the model achieves an accuracy of 93.62\%, which closely matches baseline test performance. At 25~dB and 20~dB SNR, accuracy remains stable at 93.50\% and 93.36\%, respectively. These results confirm that moderate noise has negligible impact on classification capability.

\begin{figure}[!t]
    \centering
    \centerline{\includegraphics[width=\linewidth]{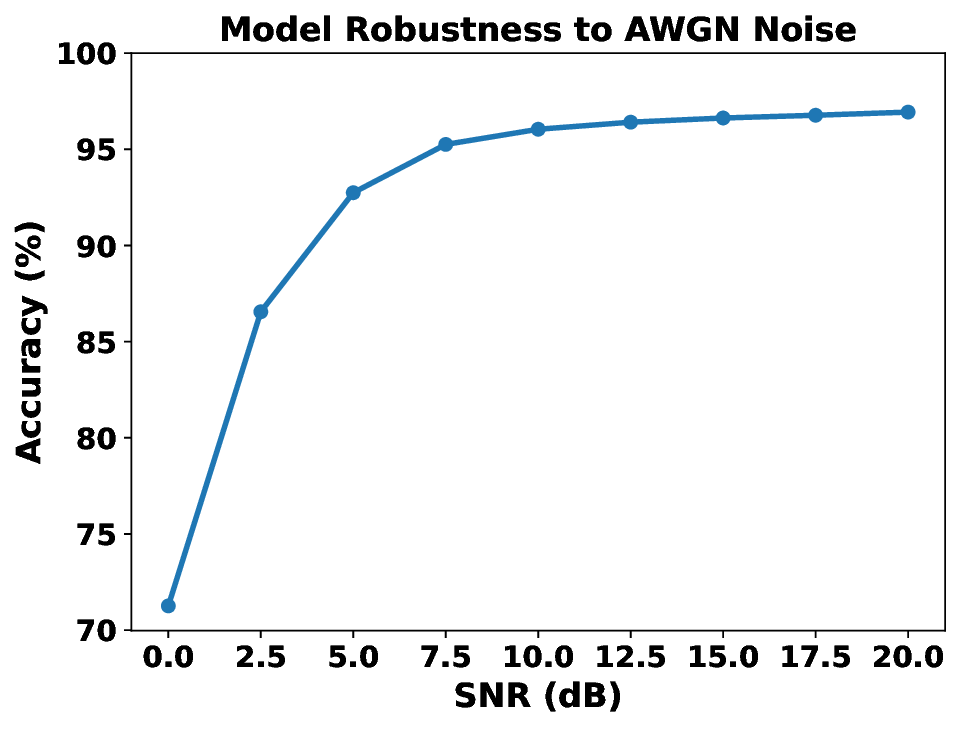}}
    \caption{Model Robustness to Noisy Inputs.}
    \label{model_snr}
\end{figure}

The effect of additive white Gaussian noise (AWGN) on classification performance was evaluated across SNR levels ranging from 0~dB to 20~dB, with the results presented in \cref{model_snr}. As the noise level increases, a gradual reduction in classification accuracy is observed. At 10~dB SNR, the model maintains an accuracy of 96.04\%, while an accuracy of 95.25\% is achieved at 7.5~dB SNR. Even at 5~dB SNR, where signal distortion becomes more pronounced, an accuracy of 92.74\% is retained. These results demonstrate that the learned feature representations preserve substantial discriminative information despite the presence of significant noise contamination.

Under more challenging conditions, strong classification performance continues to be achieved. At 2.5~dB SNR, an accuracy of 86.55\% is obtained, indicating that reliable fault identification remains possible despite considerable degradation of signal quality. At the most severe noise condition considered, namely 0~dB SNR where signal and noise powers are equal, an accuracy of 71.26\% is achieved. Although substantial corruption of the original signal is introduced under this condition, correct classification is maintained for the majority of samples.

At higher SNR levels, only minor variations in performance are observed. Accuracies of 96.41\%, 96.63\%, 96.77\%, and 96.93\% are achieved at 12.5~dB, 15~dB, 17.5~dB, and 20~dB SNR, respectively. The limited performance variation across these operating conditions indicates that the proposed model remains largely unaffected by moderate levels of additive noise.

The results presented in \cref{model_snr} indicate that performance degradation occurs progressively as the signal-to-noise ratio decreases. No abrupt decline in classification capability is observed across the evaluated SNR range. Instead, high classification accuracy is maintained throughout moderate-noise environments, while meaningful predictive capability is preserved even under severe noise exposure. Such behavior suggests that robust and noise-tolerant feature representations have been learned by the proposed architecture.

From an avionics perspective, these findings are particularly significant because sensor measurements are frequently affected by electrical interference, environmental disturbances, and measurement uncertainty. The ability of the proposed model to maintain accuracies exceeding 92\% at 5~dB SNR and 96\% at 10~dB SNR demonstrates substantial resilience to adverse operating conditions. Consequently, the obtained results support the suitability of the proposed Compact ResNet architecture for practical fault-diagnosis and health-monitoring applications in noisy operational environments.

It is noted that the original training dataset included data augmentation techniques such as additive white Gaussian noise. The SNR-based noise evaluation reported here applies additional noise during inference, on top of the training augmentation, in order to assess model behavior under explicitly controlled and increasingly adverse signal conditions. This design enables a focused evaluation of robustness beyond the noise distributions encountered during training.

\subsection{Discussion}

The experimental results demonstrate that deep learning architectures provide strong capability for multiclass detection and classification of electrical faults and power quality disturbances in aerospace power systems. All evaluated models achieve test accuracies above 94\%, which confirms the effectiveness of data-driven approaches for this application. Nevertheless, once classification performance is examined jointly with model complexity and deployment constraints, clear differences among architectures become evident.

Among the evaluated models, the 1D-CNN achieves the highest test accuracy of 97.29\%. However, this result requires more than 8.3 million parameters, which introduces substantial computational cost and memory demand. In contrast, the proposed Compact ResNet achieves a test accuracy of 96.94\% with only 175{,}685 parameters. The reduction in accuracy remains limited to 0.35 percentage points, while the decrease in model complexity exceeds an order of magnitude. Under the strict hardware, timing, and energy constraints of aerospace platforms, such a trade-off strongly favors compact architectures. These results indicate that marginal improvements in accuracy do not justify large increases in model size for embedded fault-diagnosis applications.

Further insight is provided by the comparative evaluation across architectures. Deep networks such as ResNet50 and VGG16 introduce substantial parameter growth without corresponding gains in classification performance. Similarly, hybrid architectures that combine convolutional and recurrent layers do not outperform purely convolutional models. These observations suggest that the discriminative features associated with faults and power quality disturbances are effectively captured through convolutional feature extraction without the need for additional sequential modeling. The residual connections employed in the Compact ResNet support stable training and efficient feature reuse, which enables high classification performance with a reduced number of layers and parameters.

Further evidence of the effectiveness of the proposed architecture is provided by the learning curves shown in \cref{learning_curves}. The Compact ResNet exhibits rapid convergence during the early training phase, with training accuracy exceeding 90\% within the first few epochs. Validation accuracy follows a similar trend and stabilizes near 96\% after approximately epoch 13. The gap between training and validation curves remains small throughout the training process, which indicates strong generalization and limited overfitting. The smooth convergence behavior and stability across epochs confirm that the selected architecture and training configuration enable efficient learning of discriminative features from the input data. These characteristics are particularly relevant for practical deployment, where stable and predictable training behavior supports reproducibility and reliability of the resulting model.

The class-wise analysis further supports this conclusion. The Compact ResNet achieves closely aligned precision, recall, and F1-score values, which indicates balanced classification behavior across all 21 classes. The confusion matrix shows strong concentration along the diagonal, with only limited dispersion across neighboring classes. High recall values are maintained across all classes, which reduces the likelihood of missed fault detection. This characteristic is particularly important in safety-critical aerospace systems, where reliable identification of abnormal operating conditions remains essential.

The robustness evaluation under additive white Gaussian noise provides additional evidence of the suitability of the proposed approach for real-world operating conditions. Classification accuracy decreases progressively as the signal-to-noise ratio becomes lower, without abrupt degradation in performance. At 10~dB SNR, the model maintains accuracy above 96\%, while at 5~dB SNR accuracy remains above 92\%. Even under the most severe condition considered, namely 0~dB SNR, a classification accuracy of 71.26\% is preserved. This behavior indicates that the learned representations depend on stable temporal and spectral structures rather than noise-sensitive signal amplitudes. The smooth degradation trend further confirms that the model preserves discriminative capability under significant signal corruption. The incorporation of signal-processing-based augmentation and GAN-generated samples contributes to this robustness by exposing the model to a broad range of intra-class variability during training.

The robustness results should also be interpreted within the scope of the
dataset construction protocol. The SNR study demonstrates tolerance to
controlled additive noise and supports the conclusion that the classifier uses
stable temporal and spectral class structure. However, robustness to additive
noise is not a substitute for parent-waveform-disjoint validation, since data
leakage and related-sample overlap can lead to optimistic estimates of
generalization in machine-learning studies \cite{kapoor2023leakage}. Thus,
the reported metrics provide evidence of intra-distribution performance
within the simulated aircraft-power-system operating envelope, while
independent operating-condition validation remains necessary before drawing
broader conclusions about deployment-level generalization.

The relationship between classification performance and model complexity further reinforces the advantages of the proposed architecture. The Compact ResNet occupies a favorable position in the accuracy--complexity space, where high predictive performance coincides with minimal parameter requirements. Larger models cluster in regions associated with significantly higher computational cost without clear accuracy benefits. This observation highlights the importance of efficient architectural design for practical deployment.

From a system-level perspective, the results demonstrate that lightweight architectures can achieve high classification accuracy, stable generalization, and strong robustness to noise while maintaining low computational requirements. The Compact ResNet satisfies these criteria and aligns well with the constraints of real-time aerospace systems. 

Overall, the findings confirm that carefully designed compact deep learning models provide an effective solution for fault detection and power quality monitoring in aircraft electrical systems. The balance achieved between accuracy, robustness, and efficiency supports their deployment in embedded avionics environments, where reliable operation under constrained resources remains a critical requirement.

\section{FPGA Implementation\label{sec:impl}}

The comprehensive evaluation presented in Section~\ref{sec:exp} was conducted in software and considered all candidate architectures, spanning both time--frequency-based and time-series-based learning approaches commonly applied to this problem domain; the present section focuses on FPGA deployment of the proposed model.

\subsection{Architecture Considerations for FPGA Deployment}

% The proposed Compact ResNet architecture was selected for FPGA deployment due to its balance between classification performance and architectural efficiency. Since the proposed classification pipeline relies on STFT representations, the Compact ResNet operates on two-dimensional time--frequency inputs. As reported in \cref{tab:model_comparison}, the Compact ResNet achieves a test set classification accuracy of 96.94\% with an F1 score of 93.55\% in software, while requiring only 175{,}685 parameters.

The proposed Compact ResNet architecture was selected for FPGA deployment due to its favorable balance between classification performance and architectural efficiency. Since the classification pipeline operates on STFT representations, the Compact ResNet processes two-dimensional time--frequency inputs, which enable efficient spatial feature extraction that maps naturally onto hardware. As reported in \cref{tab:model_comparison}, the Compact ResNet achieves both a test set classification accuracy and an F1 score of 96.95\% while requiring only 175{,}685 parameters. This compact model size directly reduces memory storage requirements and facilitates efficient on-chip deployment.

This compact parameter footprint represents a substantial reduction when compared with conventional two-dimensional convolutional neural network architectures commonly employed for time--frequency classification tasks. In particular, ResNet18 and ResNet50 require approximately 11.3~M and 23~M parameters, respectively, while VGG16 exceeds 138~M parameters. Even lightweight variants such as the Lightweight VGG19 architecture employ significantly larger parameter counts than the proposed Compact ResNet. Despite this reduction in model size, the Compact ResNet maintains classification performance that remains competitive with these higher-capacity models, which highlights the efficiency of the proposed architecture for STFT-based inputs.

The comparison also includes time-series-based architectures such as the 1D-CNN, 1D-CNN-LSTM, and 1D-ResNet18, which operate directly on one-dimensional signal representations. While these models demonstrate strong temporal feature extraction capability, they rely on a different representation paradigm and typically require increased architectural complexity to achieve comparable accuracy. The Compact ResNet therefore provides a balanced alternative that efficiently exploits two-dimensional time--frequency structure while maintaining a substantially reduced model footprint.

The demonstrated balance between predictive performance, robustness to noisy inputs as previously discussed, memory efficiency, and architectural compactness confirms the suitability of the Compact ResNet for FPGA-based deployment in onboard aerospace monitoring systems, where hardware resources remain limited and reliable real-time fault detection remains critical for flight safety. 

% For deployment on resource-constrained hardware, the FPGA implementation of the proposed Compact ResNet architecture was carried out with the objective of detecting electrical faults and PQDs in aircraft systems.

\subsection{Hardware Setup}
The implementation was carried out by means of the MATLAB DL HDL Toolbox, which enables the deployment of deep neural networks on FPGA devices through a dedicated DL processor IP core that supports the realization of two-dimensional convolutional neural networks (2D-CNNs) \cite{DLtoolbox}. In this work, the Xilinx Zynq UltraScale+ MPSoC ZCU102 \cite{xilinx_zcu102} development board was employed as the target platform, with Ethernet used as the interface for communication between MATLAB and the FPGA. MATLAB was executed on a desktop computer equipped with a 3.6 GHz Intel Core i7 processor, 32 GB of RAM, and a 1 TB solid-state drive (SSD) for this implementation. 
%This setup is shown in Figure (\ref{Experimental_setup}).

% \begin{figure}[!htb]
% \centerline{\includegraphics[scale=0.60]{Exp_Setup.jpg}}
% % \captionsetup{labelformat=empty}
% \caption{Experimental Setup}
% \label{Experimental_setup}
% \end{figure}

\begin{figure}
\centerline{\includegraphics[width=\linewidth]{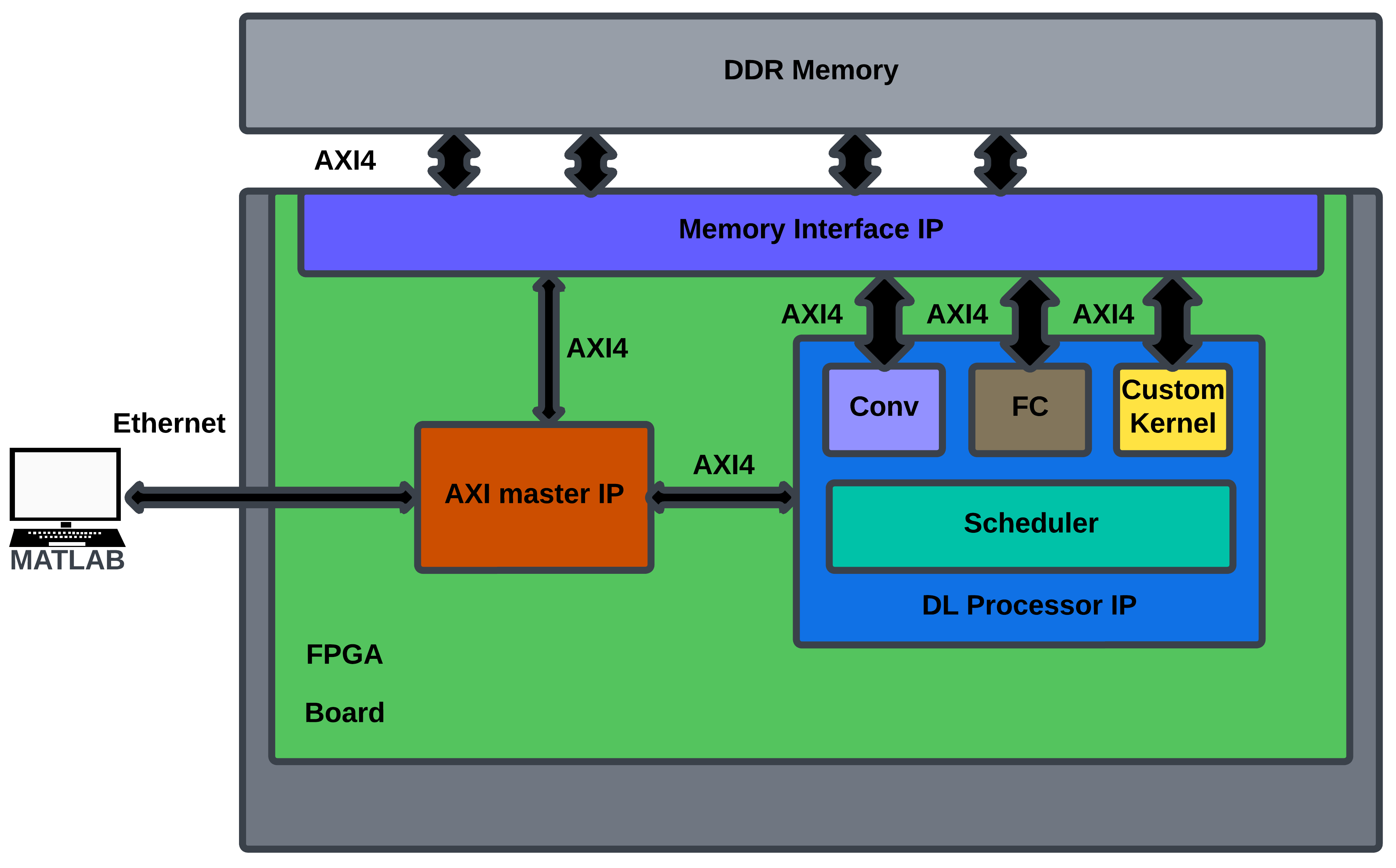}}
% \captionsetup{labelformat=empty}
\caption{MATLAB controlled DL processor.}
\label{IP_core}
\end{figure}

The general architecture of the MATLAB-controlled DL processor on FPGA is illustrated in \cref{IP_core}. An Advanced eXtensible Interface (AXI) module is employed to establish communication between the client computer and the FPGA, and the AXI4 protocol governs all data transfers among the internal components of the System-on-Chip (SoC) design \cite{Profiling_CNNs}.

\begin{figure}
    \centering
    \includegraphics[width=\linewidth]{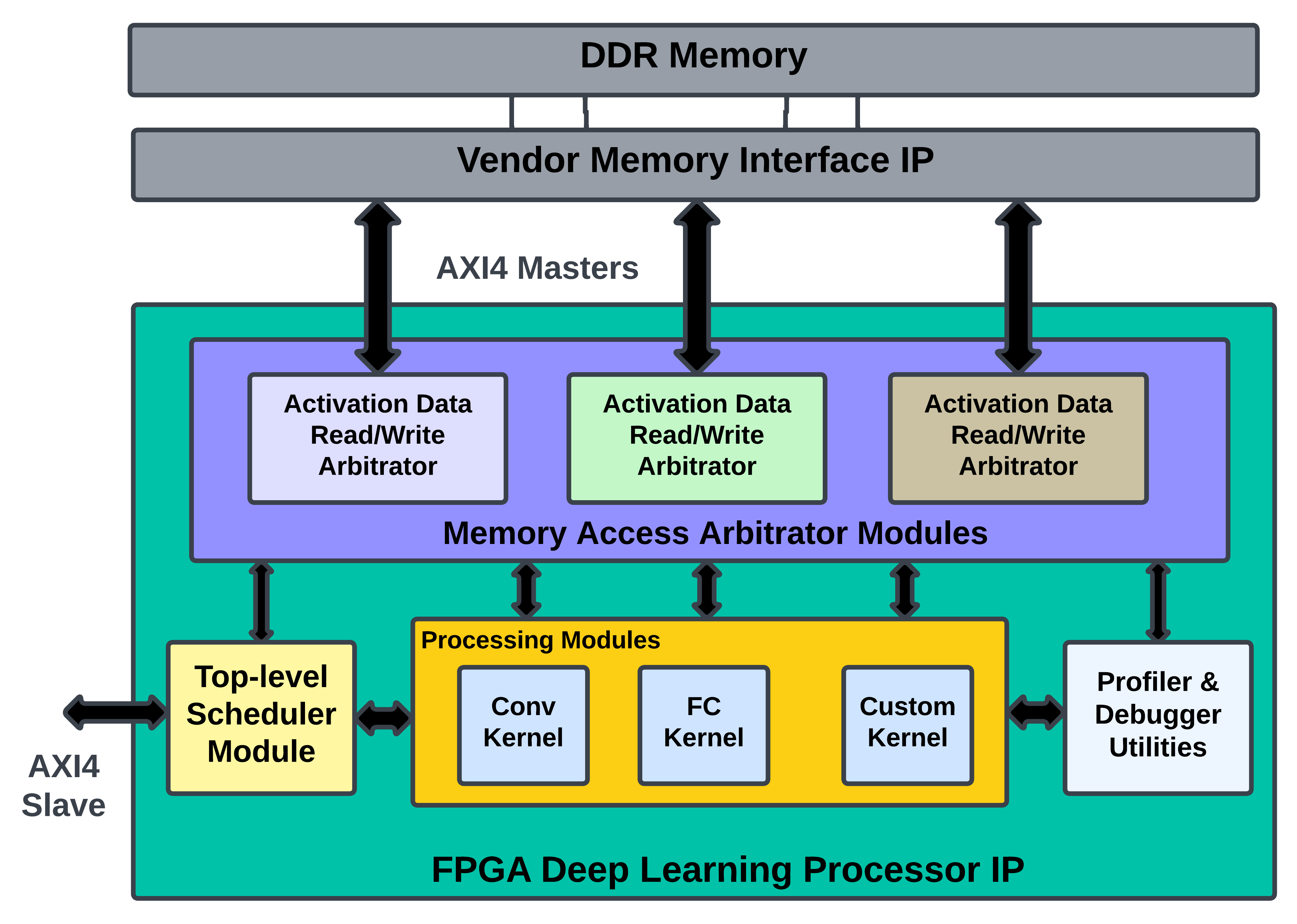} % Scale image to 80% of column width
    \caption{FPGA DL processor IP.}
    \label{DL_processor}
\end{figure}

The DL processor architecture, illustrated in \cref{DL_processor}, is composed of multiple interconnected modules, each assigned a specific function to enable efficient execution of deep neural networks on FPGA. 

External DDR memory is used to store input signals, weights, and output signals, with three AXI4 master interfaces provided for communication between the memory and the processing modules; one of these interfaces is dedicated to loading input data onto the modules. Memory access arbitrator modules are employed to control the reading and writing of weights and activation data via the AXI4 master interfaces, while a profiler interface is assigned to manage the transfer of timing data and instructions to the profiler module. The convolution kernel performs convolution operations by receiving weights and activations through dedicated AXI4 master interfaces and executing layer computations on the input signals, with support for tensors of varying sizes. Instruction execution and data retrieval from DDR memory are managed by the scheduler module, which distributes tasks to the corresponding processing modules; for example, data and commands for convolution, fully connected, and multiplication layers are forwarded to the convolution kernel, fully connected kernel, and custom kernel, respectively. The fully connected kernel retrieves weights and activations via two AXI4 master interfaces and executes fully connected layer operations, supporting tensors of different shapes and sizes. The custom kernel executes operations such as addition, multiplication, and 2D resizing layers to provide flexibility for nonstandard computations. Finally, the profiler module collects operation timing data from the kernels, including start and stop times, and generates a profiler table to support performance analysis. This modular and hierarchical design ensures efficient mapping of DL tasks onto the FPGA, providing both flexibility and high computational throughput.

\subsection{Implementation Results}

The post‑synthesis FPGA resource utilization results for the proposed deep learning processor are summarized in \cref{tab:resources_summary}. The design used 391 DSP slices out of 2520 available, which corresponds to 15.52\% of device capacity. This low DSP utilization confirms that the arithmetic workload of the neural network maps effectively to dedicated multiplier resources, thereby preserving substantial computational margin. Such margin supports avionics platforms where additional DSP capacity often serves parallel signal‑processing workloads.

\begin{table}
\centering
\caption{Post-Synthesis FPGA Resource Utilization}
\label{tab:resources_summary}
\begin{tabular}{|c|c|c|c|}
\hline
\textbf{} & \textbf{DSPs} & \textbf{Block RAM} & \textbf{CLB LUTs} \\ \hline
\textbf{Available} & 2520 & 912 & 274080 \\ \hline
\textbf{DL Processor} & 391 & 581 & 256,992  \\ \hline
\textbf{Utilization (\%)} & 15.52 & 63.71 & 93.77 \\ \hline

\end{tabular}
\end{table}

Block RAM usage reached 581 units out of 912, corresponding to 63.71\% utilization. This allocation reflects the memory requirements associated with network parameters, intermediate activations, and buffering structures. Despite this moderate use of on‑chip memory, sufficient Block RAM capacity remains available for system‑level functions such as interface buffering, redundancy mechanisms, or fault‑monitoring support that commonly arise in aerospace applications.

CLB LUT utilization reached 256,992 out of 274,080 available, which represents 93.77\% of the logic resources. Although this figure indicates dense use of the logic fabric, it also demonstrates effective consolidation of functionality within a single FPGA device. The deep learning processor integrates control logic, dataflow management, quantized arithmetic control, and memory interfacing alongside computational kernels. 

% From a system-integration perspective, the remaining LUT capacity should be interpreted cautiously. 
The present synthesis uses 93.77\% of the available CLB LUTs; therefore, the implementation demonstrates feasibility of the deep-learning accelerator itself, but it does not by itself establish that the DAQ interface, STFT computation, communication logic, and supervisory functions can all be integrated on the same FPGA without further optimization. A representative STFT architecture compatible with our DL core was shown to be implemented using only 11203 LUTs thereby fitting in the remaining area \cite{FFT_Architectures}, and most of the control and interface logic of a data acqusition system such as \cite{analog_ad7616_eval_board} can be offloaded to the ARM side of the MPSoC. Bridging the samples provided from the DAQ to the AXI stream, and buffering/FIFOs for the 32 sample STFT overlap may be possible within the remaining margin, but full integration of the acquisition and STFT subsystems requires a separate resource-budget, placement-and-routing, and timing-closure analysis which remains important future work.

The coexistence of high LUT utilization with low DSP usage and moderate Block RAM consumption suggests possible optimization pathways, such as restructuring portions of the design toward DSP-centric arithmetic, increasing reuse of memory resources, pruning the network, or applying architecture-specific HDL optimizations. Overall, the reported utilization supports accelerator-level feasibility, while complete system-level integration remains a subject for future implementation and studies.

\iffalse

From a system‑integration perspective, the remaining LUT headroom, though limited, remains sufficient for minor additions such as configuration registers, interface adaptation logic, or supervisory control blocks. In typical avionics architectures, complex auxiliary functions often reside in companion processors or software partitions, while the FPGA accelerator focuses on deterministic, high‑throughput computation. Within this context, the reported logic utilization aligns well with established integration practices in safety‑critical aerospace systems.

Furthermore, the coexistence of high LUT utilization with low DSP usage and moderate Block RAM consumption suggests possible optimization pathways, such as restructuring portions of the design toward DSP-centric arithmetic, increasing reuse of memory resources, pruning the network, or applying
architecture-specific HDL optimizations. Overall, the reported resource utilization confirms that the Compact ResNet implementation achieves efficient use of the FPGA fabric while supporting platform consolidation and deterministic operation required in avionics systems.
\fi

\subsection{FPGA Inference Performance}

The inference performance results of the proposed FPGA implementation of the DL model are summarized in \cref{fpga_perf_summary}. The reported on‑board latency of 6.90ms per sample corresponds to the execution time of the neural network accelerator on the FPGA fabric. Inference latency was measured by transmitting batches of 1000 input samples from the host computer, running MATLAB to the FPGA board and recording the total execution time for each batch. The total batch inference time was averaged across multiple runs to obtain a stable estimate, and the average per‑sample latency was then computed by dividing the mean batch execution time by the number of samples in the batch. To isolate the on‑FPGA inference latency, the round‑trip communication delay between the host and the FPGA was independently measured using a passthrough kernel and subtracted from the total measured time\cite{cooke}. As a result, the reported latency reflects on‑board computation and internal data movement. This measured latency demonstrates that the design satisfies real‑time inference requirements for embedded aerospace applications. An average latency below 10ms per sample supports rapid processing of input data while maintaining deterministic execution behavior. The corresponding throughput of 144.96 samples per second derives directly from the inverse of the average on‑board latency and reflects steady‑state operation under pipelined execution. This result confirms that the accelerator sustains continuous inference without performance degradation under sustained workloads. 

\begin{table}
\centering
\caption{Inference Performance of the FPGA Implementation}
\label{fpga_perf_summary}
\begin{tabular}{|c|c|}
\hline
\textbf{Metric} & \textbf{Value} \\ \hline
Mean On-Board Latency (ms/sample)  & 6.90   \\ \hline
Worst-Case On-Board Latency (ms/sample) & 14.40  \\ \hline
Throughput (samples/s)  & 144.96  \\ \hline
Post-Quantized Accuracy  (\%) & 95.87 \\ \hline
Accuracy Degradation (\%)  & 1.07  \\ \hline
\end{tabular}
\end{table}

Note that these latencies only reflect the latency of the DL IP core relying on a host computer to provide STFT output. A representative data acquisition system comprising a Hall-effect current sensor \cite{lem_hal50s}, voltage sensing isolation transformer \cite{lcmagnetics_iso_025kva_400hz} and a data acquisition card that integrates these sensors well with the FPGA processing unit \cite{analog_ad7616_eval_board}, as well as STFT IP cores that can process the acquired samples \cite{FFT_Architectures} would incur additional latency. While experimental verification of an integrated system's latency remains important future work, the cited representative sources claim latencies of those modules would be on the order of microseconds, three orders of magnitude less than the latency of the inference core.

The worst‑case on‑board latency was measured at 14.40ms per sample. This value corresponds to the maximum per‑sample latency observed across all classes, which occurred for class 11, and was obtained by measuring the batch inference time for 1000 samples per class and dividing by the batch size. The worst‑case on-board latency remains tightly bounded, which supports predictable scheduling and simplifies system‑level timing analysis in safety‑critical avionics systems.
In terms of classification performance, the FPGA implementation achieved a post‑quantized accuracy of 95.87\%. The floating‑point software baseline accuracy, as reported in the previous section, was 96.94\%. This difference corresponds to an absolute accuracy degradation of 1.07 percentage points. The design employed 8‑bit quantization for both weights and activations, which represents a widely adopted compromise between numerical precision and hardware efficiency in FPGA‑based neural network accelerators. The limited accuracy loss indicates that the Compact ResNet architecture exhibits robustness to reduced numerical precision and that the applied quantization strategy preserves the network’s discriminative capability.
Overall, the inference performance results demonstrate a favorable balance between latency, throughput, and accuracy. The accelerator achieves low average and worst‑case latency while maintaining classification performance close to the floating‑point baseline. When considered alongside the reported resource utilization, these results confirm that the proposed FPGA implementation meets the real‑time, deterministic, and efficiency requirements typical of avionics applications subject to strict timing and resource constraints.

\section{Concluding Remarks\label{sec:conc}}

This work investigated deep learning–based architectures implemented initially in software to explore the balance between model complexity and classification performance for the detection and classification of electrical faults and power quality disturbances in aerospace power systems. Following this evaluation, the highest‑performing architecture was selected for FPGA implementation to assess real‑time feasibility under aerospace hardware constraints. In contrast to the extensive body of literature focused on conventional 50/60Hz, this study specifically examined fault events and PQDs within aircraft electrical systems, operating at 400Hz, an area that remains largely unexplored despite the increasing prevalence of power‑electronic‑dominated architectures in MEA.
A high‑fidelity aircraft power system model inspired by the Boeing 787 electrical architecture provided the foundation for signal generation and performance evaluation. The model enabled analysis of fault and PQD behavior under high‑frequency operation and fast transient dynamics representative of aircraft environments. Compared with conventional grid‑based datasets, the generated signals exhibit distinct spectral and temporal characteristics that reflect aircraft‑specific operating conditions, thereby addressing an important gap in existing PQD research.
Evaluation of multiple deep learning architectures demonstrated that increased model complexity does not necessarily yield improved classification performance for this application. Among all evaluated models, the proposed Compact ResNet architecture achieved the most favorable trade‑off between classification accuracy and computational efficiency. The model achieved a software test accuracy of 96.94\% with only 175,685 parameters, which confirms that carefully designed lightweight architectures can meet demanding diagnostic requirements when matched appropriately to the signal representation and task structure.
The compact ResNet architecture was subsequently deployed on a Xilinx Zynq UltraScale+ MPSoC ZCU102 to evaluate hardware suitability for real‑time aerospace applications. Post-synthesis analysis confirmed efficient resource utilization, with low DSP usage, moderate Block RAM consumption, and high but acceptable logic utilization that reflects effective consolidation of control, dataflow, and arithmetic functionality within a single FPGA device. Despite the dense logic usage, LUT utilization indicates remaining headroom may support further system-level DAQ and STFT integration, validation of which remains important future work.
Inference performance results further confirm real‑time capability and deterministic behavior. The FPGA implementation achieved an average on‑board latency of 6.90ms per sample and a worst‑case latency of 14.40 ms. This tightly bounded latency supports predictable task scheduling and simplifies system‑level timing analysis, which represents a critical requirement for safety‑critical avionics certification. With 8‑bit quantization applied to weights and activations, the FPGA implementation maintained a post‑quantized accuracy of 95.87\%, corresponding to an absolute accuracy reduction of only 1.07 percentage points relative to the floating‑point software baseline. This result confirms strong resilience to reduced numerical precision and validates fixed‑point inference for embedded aircraft systems.
Robustness evaluations under additive noise conditions indicate that the proposed model preserves reliable classification performance across a wide range of signal‑to‑noise ratios. Finally, because the present validation protocol evaluates held-out stochastic
realizations rather than parent-waveform-disjoint operating campaigns, the
reported metrics should be interpreted as simulation-based
intra-distribution performance. This distinction is important because
related-sample overlap and data leakage are known to inflate apparent
machine-learning generalization performance \cite{kapoor2023leakage}.
Future work will therefore include grouped parent-waveform partitioning,
operating-condition-disjoint simulation tests, and experimental testbed
validation.

Class-level results demonstrate consistently high precision, recall, and F1-scores across all 21 classes, with low false negative rates. Such performance aligns with the safety objectives of aircraft electrical health‑monitoring systems, where reliable power system disturbance detection remains critical.
% These findings confirm the model's reliable and consistent capability to distinguish between fault and power quality disturbance classes, which is well suited for aircraft electrical health-monitoring applications.
% Class‑level analysis shows high recall for the majority of fault and PQD categories, with misclassification primarily confined to closely related disturbance types rather than nominal operating conditions. Such behavior aligns with the safety objectives of aircraft electrical health‑monitoring systems, where reliable fault detection remains critical.
Overall, this work demonstrates that compact deep learning architectures, combined with realistic aircraft power system modeling and efficient FPGA deployment, can achieve accurate, robust, and deterministic classification of electrical faults and PQDs under aerospace constraints. The presented framework establishes a foundation for intelligent onboard electrical health‑monitoring systems in MEA and highlights the feasibility of advanced PQD analysis within aircraft electrical networks. Future work will focus on integrating full DAQ/STFT/DL pipeline on the MPSoC platform, leakage-controlled experimental validation of the approach using a
ground-based aircraft-power-system testbed with real-flight data, extending classification to additional fault types, and eventual integration of the system into onboard health-monitoring platforms to enhance reliability and predictive maintenance in next-generation aircraft.

\section*{ACKNOWLEDGMENT}
Text portions of this manuscript were refined using Microsoft 365 Copilot Writing Coach Agent. The final content was thoroughly reviewed and approved by the authors.

\bibliographystyle{jabbrv_IEEEtran}
\bibliography{IEEEabrv,References}

% \begin{IEEEbiography}

\begin{IEEEbiography}[{%
  \includegraphics[
    width=1in,
    height=1.25in,
    clip,
    keepaspectratio
  ]{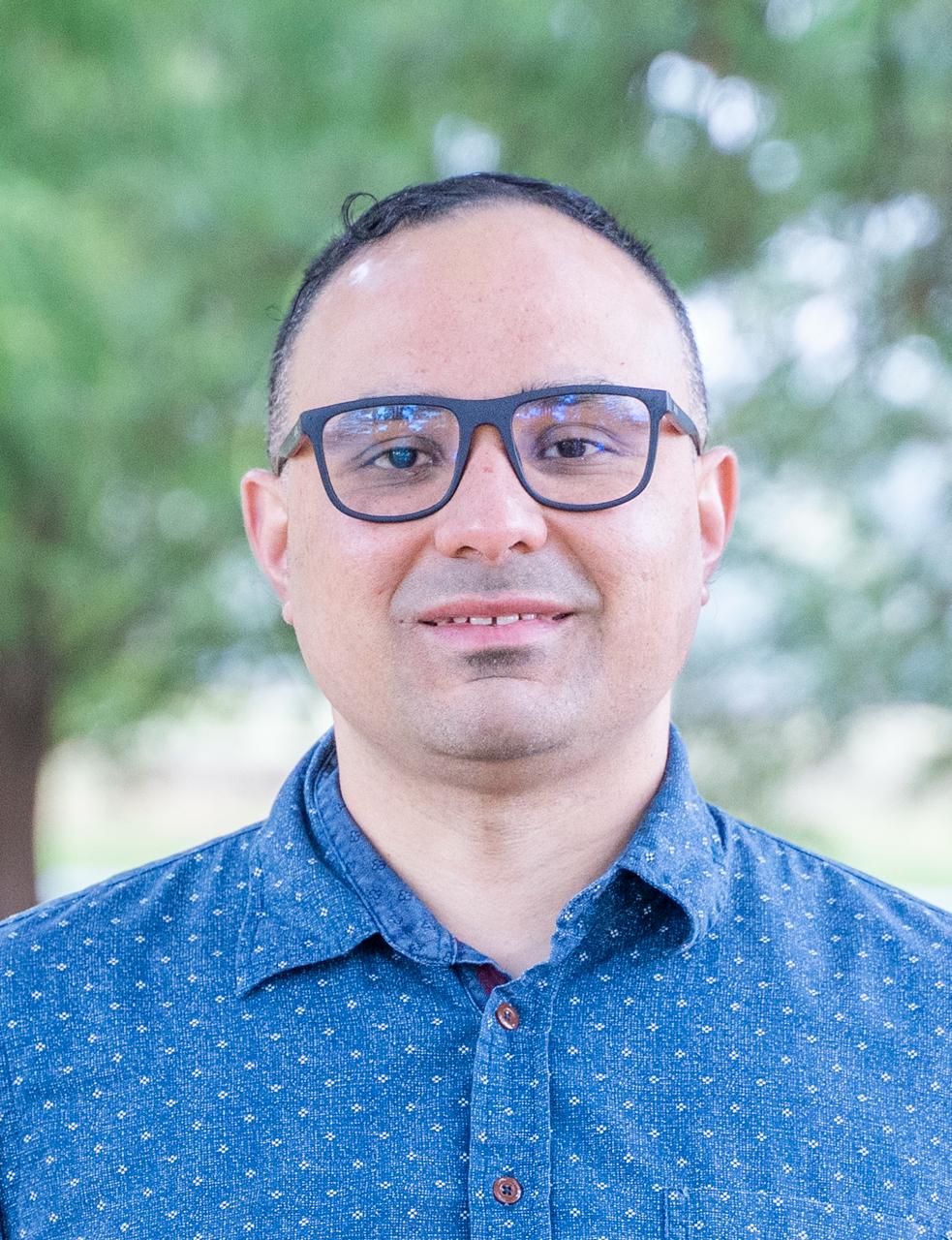}% 
}]{Ian C. Guzmán} received the B.S. degree in electronic engineering and the M.S. degree in electrical engineering from Universidad del Valle, Cali, Colombia, in 2013 and 2017, respectively. He received the M.S. degree in electrical and computer engineering from the University of Delaware, Newark, DE, USA, in 2022, and the Ph.D. degree in electrical engineering and computer science from Embry-Riddle Aeronautical University, Daytona Beach, FL, USA, in 2026.

From 2020 to 2022, he was a Research Assistant with the University of Delaware. From 2023 to 2026, he was a Research Assistant with Embry-Riddle Aeronautical University. He has served as an Adjunct Professor of computer science at La Salle University, Philadelphia, PA, USA, and Gwynedd Mercy University, Gwynedd Valley, PA, USA. He is currently an Adjunct Professor of computer science at La Salle University, where he teaches undergraduate computer science and graduate-level data science courses.

His research interests span the intersection of data science, machine learning, deep learning, digital signal processing, power systems, and FPGA-based computing.

\end{IEEEbiography}

\begin{IEEEbiography}[{%
  \includegraphics[
    width=1in,
    height=1.25in,
    clip,
    keepaspectratio
  ]{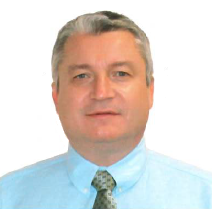}%
}]{Radu F. Babiceanu} (Senior Member, IEEE) earned his degree in industrial and systems engineering from Virginia Tech in 2005. From January 2025, he is a Professor and the Chair of the Department of Electrical and Computer Engineering at Western Michigan University. Starting with the Fall 2026 semester, he also serves as the Interim Chair of the Computer Science Department at Western Michigan. Previously, he was on the faculty and served as Interim Chair of the Electrical Engineering and Computer Science Department at Embry-Riddle Aeronautical University in Daytona Beach, FL.

Throughout his faculty years, he developed and taught a large number of fundamental and advanced systems engineering courses with an overarching aviation ecosystem theme. He also developed and delivered training courses for industry in the areas of aircraft safety engineering and certification, and aviation cybersecurity. His main research interests are in the aviation and aerospace operational ecosystem, with a focus in cybersecurity and safety-critical systems assurance, cyber-, engineering-, operational-, and organizational-resilience, and AI/ML approaches to enhanced operations.

Dr. Babiceanu's professional service includes serving as Associate and Guest Editor, and on the Editorial Board of well-known scientific journals, conference organizer and program chair, leadership positions with ASEE and INCOSE, and panel reviewer for scholarly research and innovation competitions. He is a senior member of IISE.

\end{IEEEbiography}

\begin{IEEEbiography}[{%
  \includegraphics[
    width=1in,
    height=1.25in,
    clip,
    keepaspectratio
  ]{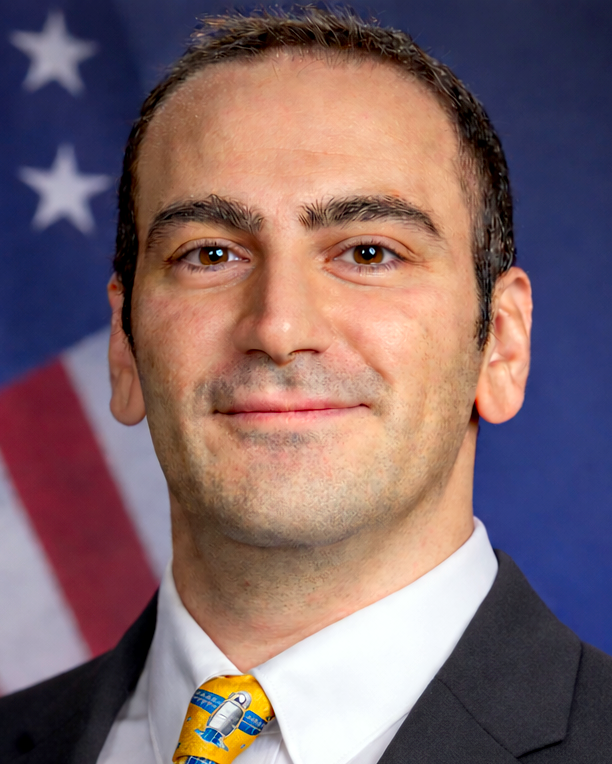}%
}]{Berker Peköz} (Member, IEEE) received the B.S. degree in electrical and electronics engineering from Middle East Technical University, Ankara, Türkiye, in 2015, and the M.S.E.E. and Ph.D. degree in electrical engineering from the University of South Florida, Tampa, FL, USA, in 2017 and 2020, respectively. 

From 2020 to 2023, he was a Senior Wireless DSP/ASIC Systems Engineer with Qualcomm Technologies, Inc., Bridgewater, NJ, USA. He joined CesiumAstro, Inc., Austin, TX, USA, in 2023, as a Senior Communications Systems Engineer, and later served as a Scientific Consultant. He was a NASA Glenn Faculty Fellow with the NASA Glenn Research Center, Cleveland, OH, USA, in 2026. Since 2025, he has served as the
Scientific Advisor (a member of the Board of Advisors) of Kenyi Technologies, Inc., Newark, NJ, USA. He is currently an Assistant Professor of electrical engineering and computer science at Embry-Riddle Aeronautical University, Daytona Beach, FL, USA.

Dr. Peköz is an inventor on eight U.S. patents, a member of the National Academy of Inventors, Tau Beta Pi, and COnsortium for Space Mobility and ISAM Capabilities (COSMIC) Academia Caucus. He has
served as a technical reviewer for several IEEE Transactions and Magazines and on the Technical Program Committees for several IEEE
conferences and workshops, such as the IEEE/IFIP International Conference on Dependable Systems and Networks
(DSN) Workshop on Dependable
and Secure Autonomous Systems (DSAS). For more information, see \url{https://faculty.erau.edu/Berker.Pekoz}.

\end{IEEEbiography}

\end{document}